\documentclass[preprint,aps,12pt,showpacs,nofootinbib,tightenlines]{revtex4}
\usepackage{amsmath}
\usepackage{amssymb}
\usepackage{epsfig}
\usepackage{graphicx}
\begin{document}

\newcommand{\beq}{\begin{eqnarray}}
\newcommand{\eeq}{\end{eqnarray}}
\newcommand{\non}{\nonumber\\ }

\newcommand{\etar}{\eta^\prime }
\newcommand{\etap}{\eta^{(\prime)} }
\newcommand{\acp}{ {\cal A}_{CP} }
\newcommand{\cala}{ {\cal A} }
\newcommand{\mw}{ M_W}

\newcommand{\pka}{\phi_{K}^A}
\newcommand{\pkp}{\phi_K^P}
\newcommand{\pkt}{\phi_K^T}

\newcommand{\pea}{\phi_{\eta}^A}
\newcommand{\pep}{\phi_{\eta}^P}
\newcommand{\pet}{\phi_{\eta}^T}
\newcommand{\peqa}{\phi_{\eta_q}^A}
\newcommand{\peqp}{\phi_{\eta_q}^P}
\newcommand{\peqt}{\phi_{\eta_q}^T}

\newcommand{\pesa}{\phi_{\eta_s}^A}
\newcommand{\pesp}{\phi_{\eta_s}^P}
\newcommand{\pest}{\phi_{\eta_s}^T}

\newcommand{\pepa}{\phi_{\eta'}^A}
\newcommand{\pepp}{\phi_{\eta'}^P}
\newcommand{\pept}{\phi_{\eta'}^T}

\newcommand{\pksa}{\phi_{K^*}}
\newcommand{\pksp}{\phi_{K^*}^s}
\newcommand{\pkst}{\phi_{K^*}^t}

\newcommand{\fb}{f_B }
\newcommand{\fk}{f_K }
\newcommand{\fe}{f_{\eta} }
\newcommand{\fep}{f_{\eta'} }
\newcommand{\rks}{r_{K^*} }
\newcommand{\rk}{r_K }
\newcommand{\re}{r_{\eta} }
\newcommand{\rep}{r_{\eta'} }
\newcommand{\mb}{m_B }

\newcommand{\xeba}{\bar{x}_2}
\newcommand{\xsba}{\bar{x}_3}
\newcommand{\res}{r_{\eta_s}}
\newcommand{\red}{r_{\eta_q}}
\newcommand{\peas}{\phi^A_{\eta_s}}
\newcommand{\peps}{\phi^P_{\eta_s}}
\newcommand{\pets}{\phi^T_{\eta_s}}
\newcommand{\pead}{\phi^A_{\eta_q}}
\newcommand{\pepd}{\phi^P_{\eta_q}}
\newcommand{\petd}{\phi^T_{\eta_q}}

\newcommand{\pvsl}{ p \hspace{-2.0truemm}/_{K^*} }
\newcommand{\esl}{ \epsilon \hspace{-2.1truemm}/ }
\newcommand{\psl}{ P \hspace{-2.4truemm}/ }
\newcommand{\nsl}{ n \hspace{-2.2truemm}/ }
\newcommand{\vsl}{ v \hspace{-2.2truemm}/ }
\newcommand{\epsl}{\epsilon \hspace{-1.8truemm}/\,  }
\newcommand{\bfkk}{{\bf k} }
\newcommand{\calm}{ {\cal M} }
\newcommand{\calh}{ {\cal H} }
\newcommand{\ov}{ \overline  }

\def \cpl{ Chin. Phys. Lett.  }
\def \ctp{ Commun. Theor. Phys.  }
\def \epjc{ Eur. Phys. J. C }
\def \jpg{  J. Phys. G }
\def \npb{  Nucl. Phys. B }
\def \plb{  Phys. Lett. B }
\def \prd{  Phys. Rev. D }
\def \prl{  Phys. Rev. Lett.  }
\def \zpc{  Z. Phys. C }
\def \jhep{ J. High Energy Phys.  }

\title{$B_{s} \to PP $ decays and the NLO contributions in the pQCD Approach}
\author{Jing Liu, Rui Zhou and Zhen-Jun Xiao
\footnote{xiaozhenjun@njnu.edu.cn},}
\affiliation{Department of Physics and Institute of
Theoretical Physics, Nanjing Normal University, Nanjing, Jiangsu 210097, P.R.China }
\date{\today}
\begin{abstract}
By employing the perturbative QCD(pQCD) factorization approach, we calculated the
partial next-to-leading order (NLO) contributions to
$B_s \to P P$ decays ( $P= \pi, K, \etap $ ), coming from the QCD vertex corrections,
the quark-loops and the chromo-magnetic penguins.
we found numerically that
(a) for three measured decays $\bar{B}_s \to K^+ \pi^-, K^+K^-$ and $\pi^+\pi^-$,
the consistency between the pQCD predictions
and the measured values are improved effectively by the inclusion of the NLO
contributions;
(b) for $\bar{B}_s \to K^0\etap$ and $K^0\pi^0$ decays,
the NLO enhancements to the branching ratios can be significant, from $\sim 50\%$ to $170\%$,
to be tested by the LHC experiments;
(c) for the CP-violating asymmetries, the leading order pQCD
predictions can also be changed significantly by the inclusion of the NLO contributions;
(d) for $\bar{B}_s \to K^+\pi^-$ decay, the pQCD prediction for the
direct CP asymmetry is $\acp^{dir}(\bar{B}_s \to K^+\pi^-)=0.26\pm 0.06$, which
agrees very well with the only measured value available currently.

\end{abstract}

\vspace{1cm}


\maketitle

\section{Introduction}

The two-body charmless hadronic decays of $B$ or $B_s$ meson are the good place to test
the Standard Model (SM) and look for the signal of new physics beyond the SM.
Since 1999, more than $10^9$ events of $B \bar{B}$ pair production and decay have been
collected and studied in the B factory experiments.
In the Large Hadron Collider (LHC) experiments (ATLAS, CMS and LHC-b),
besides those light  $B_{u,d}$ mesons, a huge number of heavier $B_s$ meson
production and decay events will be collected \cite{lhcb1}.
The study about the charmless decays of $B_s$ meson
is therefore becoming more interesting then ever before.

By employing the generalized factorization approach\cite{aag,yhc} or the
QCD factorization (QCDF) approach \cite{bbns99}, about 40 $B_s \to M_2 M_3$
($M_i$ stands for the light pseudo-scalar or vector mesons ) decay modes
have been studied,  for example, in the framework of SM \cite{chenbs99,npb675} or
in some new physics models beyond the SM \cite{xiaobs01}.
Many $B_s$ meson decays have also been calculated, on the other hand,
by employing the perturbative QCD (pQCD) factorization approach at
leading order \cite{pipi,pieta,ali07}.

Very recently, some next-to-leading order (NLO) contributions to some $B \to M_2 M_3$ decays
have been calculated by employing the pQCD approach\cite{nlo05,xiao08a,xiao08b}.
One can see from those studies that the NLO contributions can change significantly the
leading order (LO) pQCD predictions for some decay modes. It is therefore necessary
to calculate the NLO contributions to those two-body charmless $B_s$ meson
decays, in order to improve the reliability of the theoretical predictions.

we here focus on the calculations of NLO contributions to $B_s \to P P$ decays
( $P= \pi, K, \etap $ ) in the pQCD approach. The NLO contributions considered here
include: QCD vertex corrections,
the quark-loops and the chromo-magnetic penguins.
We expect that they are the major part of the full NLO contributions in pQCD
approach \cite{nlo05}.
The remaining NLO contributions in pQCD approach, such as those from
the factorizable emission diagrams, hard-spectator and annihilation diagrams as illustrated
in Figs.~5-7 in Ref.~\cite{xiao08b}, have not been calculated at present and
should be studied as soon as possible.

This paper is organized as follows. In Sec.~\ref{sec:f-work}, we
give a brief review about the pQCD factorization approach.
In Sec.~\ref{sec:lo}, we calculate analytically the relevant Feynman diagrams and
present the various decay amplitudes for the studied decay modes in
the leading-order.
In Sec.~\ref{sec:nlo}, the NLO contributions from the vertex corrections, the
quark loops and the chromo-magnetic penguin amplitudes are evaluated.
We calculate and show the pQCD predictions for the branching ratios and  CP violating
asymmetries of $B_s \to PP$ decays in Sec.~V.
The summary and some discussions are included in the final section.

\section{ Theoretical framework}\label{sec:f-work}

\subsection{ Decay amplitude in pQCD}\label{sec:2-1}

In the pQCD approach, the decay amplitude is separated into soft ($\Phi_{M_i}$), hard
( $H(k_i,t)$ ), and harder( $C(M_W)$ ) dynamics characterized by different energy
scales $( \Lambda_{QCD}, t, m_b, M_W )$ \cite{li2003}.
The decay amplitude ${\cal A}(B \to M_2 M_3)$ can be written conceptually as the convolution,
\beq
{\cal A}(B \to M_2 M_3)\sim \int\!\! d^4k_1 d^4k_2 d^4k_3\ \mathrm{Tr}
\left [ C(t) \Phi_B(k_1) \Phi_{M_2}(k_2) \Phi_{M_3}(k_3)
H(k_1,k_2,k_3, t) \right ],
\label{eq:con1}
\eeq
where $k_i$'s are momenta of light quarks included in each meson, and $\mathrm{Tr}$
denotes the trace over Dirac and color indices. $C(t)$ is the Wilson
coefficient evaluated at scale $t$.
The hard function $H(k_1,k_2,k_3,t)$ describes the four quark operator and the
spectator quark connected by  a hard gluon whose $q^2$ is in the order
of $\bar{\Lambda} M_B$, and can be perturbatively calculated.
The function $\Phi_{M_i}$ is the wave function which describes hadronization of the
quark and anti-quark in the meson $M_i$.
While the hard kernel $H$ depends on the processes considered,
the wave function $\Phi_{M_i}$ is independent of the specific processes.
Using the wave functions determined from other well measured processes, one can make
quantitative predictions here.

Since the b quark inside the B meson is rather heavy, we consider the $B$ meson at rest
for simplicity.
Using the light-cone coordinates, we define the emitted meson $M_2$ moving along
the direction of $n=(1,0,{\bf 0}_{\rm T})$ and the recoiled meson $M_3$ the direction of
$v=(0,1,{\bf 0}_{\rm T})$. Here we also use $x_i$ to denote the momentum fraction
of anti-quark in each meson:
\beq
P_{B_s} &=& \frac{M_B}{\sqrt{2}} (1,1,{\bf 0}_{\rm T}), \quad
P_2 = \frac{M_{B_s}}{\sqrt{2}}(1,0,{\bf 0}_{\rm T}), \quad
P_3 = \frac{M_{B_s}}{\sqrt{2}} (0,1,{\bf 0}_{\rm T}),\non
k_1 &=& (x_1 P_1^+,0,{\bf k}_{\rm 1T}), \quad
k_2 = (x_2 P_2^+,0,{\bf k}_{\rm 2T}), \quad
k_3 = (0, x_3 P_3^-,{\bf k}_{\rm 3T}).
\eeq
Then, the integration over $k_1^-$, $k_2^-$, and $k_3^+$ in
eq.(\ref{eq:con1}) will lead to
\beq
{\cal A} &\sim
&\int\!\! d x_1 d x_2 d x_3 b_1 d b_1 b_2 d b_2 b_3 d b_3 \non &&
\cdot \mathrm{Tr} \left [ C(t) \Phi_B(x_1,b_1) \Phi_{M_2}(x_2,b_2)
\Phi_{M_3}(x_3, b_3) H(x_i, b_i, t) S_t(x_i)\, e^{-S(t)} \right ],
\quad \label{eq:a2}
\eeq
where $b_i$ is the conjugate space coordinate of $k_{iT}$.
The large logarithms ($\ln m_W/t$) coming
from QCD radiative corrections to four quark operators are included
in the Wilson coefficients $C(t)$. The large double logarithms
($\ln^2 x_i$) on the longitudinal direction are summed by the
threshold resummation, and they lead to $S_t(x_i)$ which
smears the end-point singularities on $x_i$. The last term,
$e^{-S(t)}$, is the Sudakov form factor which suppresses the soft
dynamics effectively \cite{li2003}.

\subsection{ Effective Hamiltonian and Wilson coefficients}\label{sec:2-1b}

For the studied $B_s \to PP$ decays, the weak effective Hamiltonian $H_{eff}$
for $b \to s$ transition can be written as \cite{buras96}
\beq
\label{eq:heff}
{\cal H}_{eff} = \frac{G_{F}}
{\sqrt{2}} \, \sum_{q=u,c}V_{qb} V_{qs}^*\left\{  \left [ C_1(\mu)
O_1^q(\mu) + C_2(\mu) O_2^q(\mu) \right ]
+ \sum_{i=3}^{10} C_{i}(\mu) \;O_i(\mu) \right\} \; .
\eeq
where $G_{F}=1.166 39\times 10^{-5} GeV^{-2}$ is the Fermi constant,
and $V_{ij}$ is the Cabbibo-Kobayashi-Maskawa (CKM) matrix element,
$C_i(\mu)$ are the Wilson coefficients evaluated
at the renormalization scale $\mu$ and $O_i(\mu)$ are the four-fermion operators.
For the case of $b \to d $ transition, simply makes a replacement of $s$ by
$d$ in Eq.~(\ref{eq:heff}) and in the
expressions of $O_i(\mu)$ operators, which can be found easily for example in
Ref.\cite{buras96}.

In PQCD approach, the energy scale $``t"$ is chosen as the largest energy scale in
the hard kernel $H(x_i,b_i,t)$ of a given Feynman diagram, in order to
suppress the higher order corrections and improve the reliability of the perturbative
calculation.
Here, the scale $``t"$ may be larger or smaller than the $m_b$ scale.
In the range of $t \geq m_b$, the number of active quarks is $N_f=5$, and
the renormalization group (RG) running of  the Wilson
coefficients $C_i(t)$ and LO and NLO level can be written as \cite{buras96}.
\beq
C_i(t)^{LO} &=&  U(t,\mw)^{(0)}_{ij} C_j(\mw)^{LO}, \non
C_i(t)^{NLO} &=& U(t,\mw,\alpha)_{ij} C_{j}(\mw)^{NLO}.
\label{eq:cit-01}
\eeq
The explicit expressions of $C_i^{LO, NLO}(\mw)$, the RG evolution matrix $U(t,\mw)^{(0)}$ and
$U(t,\mw,\alpha)$ can be found easily, for example, in
Refs.~\cite{buras96}.

In the range of $ \mu_0 \leq  t < m_b $, the number of active quarks is $N_f=4$,
and we have similarly
\beq
C_i(t)^{LO} &=&  U(t,m_b)^{(0)}_{ij} C_j(m_b)^{LO}, \non
C_i(t)^{NLO} &=& U(t,m_b,\alpha)_{ij} C_{j}(m_b)^{NLO}.
\label{eq:cit-02}
\eeq

According to the analysis in Ref.~\cite{xiao08b}, we believe that it is reasonable to
choose $\mu_0=1.0$ GeV as the lower cut-off of the hard scale $t$, which is also close
to the hard-collinear scale $\sqrt{\bar{\Lambda}m_B} \sim 1.3$ GeV in SCET \cite{scet01}.
In the numerical integrations we will
fix the values $C_{i}(t)$ at $C_{i}(1.0)$  whenever the scale $t$ runs below the
scale $\mu_0=1.0$ GeV.

\subsection{ Wave functions}\label{sec:2-3}

As usual, we treat the $B$ meson as a very good heavy-light system, and
consider only the contribution of Lorentz structure
\beq
\Phi_{B_s}= \frac{1}{\sqrt{2N_c}}
(\psl_{B_s} +m_{B_s}) \gamma_5 \phi_{B_s} ({\bf k_1}),
\label{eq:bsmeson}
\eeq
with
\beq
\phi_{B_s}(x,b)&=& N_{B_s} x^2(1-x)^2 \mathrm{exp} \left
 [ -\frac{M_{B_s}^2\ x^2}{2 \omega_{b}^2} -\frac{1}{2} (\omega_{b} b)^2\right],
 \label{phib}
\eeq
where $\omega_{b}$ is a free parameter and we take
$\omega_{b}=0.5\pm 0.05$ GeV for $B_s$ meson. For a given $\omega_b$,
the normalization factor $N_{B_s}$ can be determined through the normalization
condition
\beq
\int\frac{d^4 k_1}{(2\pi)^4}\phi_{B_s}({\bf k_1})
=\frac{f_{B_s}}{2\sqrt{6}},
\eeq
with $f_{B_s}=230$ MeV.

For the $\eta-\eta^\prime$ system, we employ  the quark-flavor mixing scheme:
the physical states $\eta$ and $\etar$ are related to the flavor states
$\eta_q= (u\bar u +d\bar d)/\sqrt{2}$  and $\eta_s=s\bar{s}$
through a single mixing angle $\phi$,
\beq
\left(\begin{array}{c} \eta \\ \eta^{\prime} \end{array} \right)
=\left(\begin{array}{cc}
 \cos{\phi} & -\sin{\phi} \\
 \sin{\phi} & \cos{\phi} \\ \end{array} \right)
 \left(\begin{array}{c} \eta_q \\ \eta_s \end{array} \right).
\label{eq:e-ep}
\eeq
The relation between the decay constants $(f_\eta^q, f_\eta^s,f_{\etar}^q,f_{\etar}^s)$ and
$(f_q,f_s,)$, the chiral enhancement $m_0^q$ and $m_0^s$ associated
with the two-parton twist-3 $\eta_q$ and $\eta_s$ meson
distribution amplitudes (DA's) have been defined in Ref.\cite{nlo05}.
The three relevant input parameters $f_q, f_s$ and $\phi$ have been
extracted from the data of the relevant exclusive processes \cite{fks98}:
\beq
f_q=(1.07\pm 0.02)f_{\pi},\quad f_s=(1.34\pm 0.06)f_{\pi},\quad \phi=39.3^\circ\pm 1.0^\circ,
\eeq
with $f_\pi=130$ MeV.

For the light pseudo-scalar mesons $\pi$ and $K$, as well as $\eta_q$ and $\eta_s$,
their wave functions are the same in form and can be defined as
\cite{pball98}
\beq
\Phi(P,x,\zeta)\equiv \frac{1}{\sqrt{2N_C}}\gamma_5 \left [ \psl
\phi^{A}(x)+m_0 \phi^{P}(x)+ \zeta m_0 (\nsl \vsl -1)\phi_{P}^{T}(x)\right ],
\label{eq:phi-x1}
\eeq
where $P$ and $x$ are the momentum of the light meson and the momentum fraction of the quark
(or anti-quark) inside the meson, respectively.
When the momentum fraction of the quark (anti-quark) is set to be $x$, the parameter
$\zeta$ should be chosen as $+1$ ($-1$).

The expressions of the relevant DA's of the meson $M=(\pi, K, \eta_q, \eta_s)$
are the following \cite{pball98}:
\beq
\phi_M^A(x) &=&  \frac{3 f_M}{\sqrt{6} } x (1-x)
    \left[1+a_1^{M}C^{3/2}_1(t)+a^{M}_2C^{3/2}_2(t)\right],\label{eq:piw1}\\
\phi_M^P(x) &=&   \frac{f_M}{2\sqrt{6} }
   \left [ 1+\left (30\eta_3-\frac{5}{2}\rho^2_{M} \right ) C^{1/2}_2(t)
   \right ], \ \
\label{eq:piw2}   \\
\phi_M^T(x) &=&  \frac{f_M(1-2x)}{2\sqrt{6} }
   \left[ 1+6\left (5\eta_3-\frac{1}{2}\eta_3\omega_3-\frac{7}{20}\rho^2_M
   -\frac{3}{5}\rho^2_M a_2^{M} \right )
   \left (1-10x+10x^2\right )\right],\quad
   \label{eq:piw3}
\eeq
with the mass ratio $\rho_M=(m_\pi/m_0^\pi, m_K/m_0^K,m_{qq}/m_0^q,m_{ss}/m_0^s)$ for
$M=(\pi, K, \eta_q, \eta_s)$ respectively \cite{nlo05,xiao08b}.
The Gegenbauer moments $a_i^M$ have been chosen as \cite{ali07}:
\beq
a^\pi_1&=&a_1^{\eta_q,\eta_s}= 0,\quad a^\pi_2=a_2^{\eta_q,\eta_s}=0.44,
\non
a^K_1  &=& 0.17,\quad a^K_2=0.20.
\eeq
The values of other parameters are $\eta_3=0.015$ and $\omega=-3.0$.
At last the Gegenbauer polynomials $C^{\nu}_n(t)$ in Eqs.~(\ref{eq:piw1}-\ref{eq:piw3})
are defined as:
\beq
C^{3/2}_1(t)=3t,\quad
C^{1/2}_2(t)=\frac{1}{2}(3t^2-1), \quad
C^{3/2}_2(t)=\frac{3}{2}(5t^2-1),
\label{eq:c124}
\eeq
with $t=2x-1$. Like Ref.~\cite{ali07}, we here also drop the terms proportional to
$C_4^{1/2,3/2}(t)$ in the LCDA's for $\phi_\pi^{A}$, $\phi_\pi^P$, and $\phi_K^T$.

There are many studies about the distribution amplitudes of light mesons
and the relevant Gagenbauer moments \cite{pball98}.
In recent years, the light-cone distribution amplitudes have been updated
continually\cite{pball06}. The inclusion of higher order terms and
the variations of the Gegenbauer moments do affect the
theoretical predictions for the branching ratios and CP-violating asymmetries
of the considered decays, but the resultant changes of theoretical predictions
are indeed not significant, according to the analysis in Ref.~\cite{ali07}.


\section{ Decay amplitudes at leading order}\label{sec:lo}

The thirteen $B_s^0 \to PP$ decays ($P=\pi, K, \eta, \etar$) have been studied previously in
Ref.~\cite{ali07} by employing the pQCD factorization approach at leading order.
The decay amplitudes as presented in Ref.\cite{ali07} are confirmed
by our recalculation. In this paper, we focus on
the calculations of some NLO contributions to these decays in the pQCD factorization approach.

\begin{figure}[tb]
\vspace{-5cm}
\centerline{\epsfxsize=18cm \epsffile{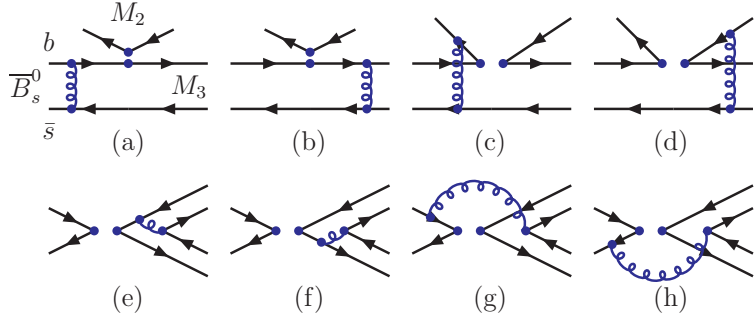}}
\vspace{-16cm}
\caption{ Feynman diagrams which may contribute at leading order
to $B_s^0 \to PP$ decays.}
 \label{fig:fig1}
\end{figure}

At the leading order, the relevant Feynman diagrams
which may contribute to the $B_s^0\to PP$ decays are  illustrated in Fig.~\ref{fig:fig1}.
For the sake of completeness, however, we firstly show the  relevant
LO decay amplitudes in this section based on our own analytical calculations.

For $\bar {B}_s^0\to K^0 \etap$ decays, the LO decay amplitudes are
\beq
{\cal A}({\bar B}_s^0 \to K^0 \eta )&=&
{\cal A}({\bar B}_s^0\to K^0 \eta_q )\cos(\phi)
-{\cal A}({\bar B}_s^0\to K^0\eta_s)\sin(\phi),\label{eq:aeta}
\\
{\cal A}({\bar B}_s^0\to K^0 \etar ) &=&
{\cal A}({\bar B}_s^0\to K^0 \eta_q )\sin(\phi)+{\cal A}({\bar B}_s^0\to K^0\eta_s)\cos(\phi),
\label{eq:aetap}
\eeq
with
\beq
\sqrt{2}{\cal A}(\bar {B}_s^0\to K^0 \eta_q) &=&
\xi_u \left ( f_q F_{eK}\; a_2+ M_{eK} \; C_2 \right ) \non
&&
- \xi_t \left \{ f_q F_{eK} \left ( 2a_3+a_4-2a_5-\frac{1}{2}a_7+\frac{1}{2}a_9
-\frac{1}{2}a_{10} \right )
\right. \non
&& \left.
+M_{eK}\left (C_3+2C_4+2C_6+\frac{1}{2}C_8-\frac{1}{2}C_9
+\frac{1}{2}C_{10}\right )
\right. \non
&& \left.
+ f_{B_s} F_{aK} \left ( a_4-\frac{1}{2}a_{10} \right )
+\left(f_{B_s} F_{aK}^{P_2} + f_q F_{eK}^{P_2}\right)
\left (a_6-\frac{1}{2}a_8 \right )
\right. \non
&& \left.
+ M_{aK} \left (C_3-\frac{1}{2}C_9 \right ) +
M_{aK}^{P_1} \left (C_5-\frac{1}{2}C_7 \right) \right \},
\label{eq:k0etaq}
\eeq
\beq
{\cal A}({\bar B}_s^0\to K^0\eta_s) &=&
-\xi_t \left \{ f_s F_{eK} \left ( a_3-a_5+\frac{1}{2}a_7-\frac{1}{2}a_9 \right)
\right. \non
&& \left.
+\left ( f_K F_{e\eta_s} + + f_{B_s} F_{aK}\right)
\left ( a_4-\frac{1}{2}a_{10} \right )
\right. \non
&& \left.
+ \left (f_K F_{e\eta_s}^{P_2} + f_{B_s} F_{aK}^{P_2} \right )
\left (a_6-\frac{1}{2}a_8 \right)
+M_{eK}\left (C_4+C_6-\frac{1}{2}C_8-\frac{1}{2}C_{10} \right)
\right. \non
&& \left.
+ \left (M_{e\eta_s} + M_{aK}\right )\left (C_3-\frac{1}{2}C_9 \right)
+ \left ( M_{e\eta_s}^{P_1} + M_{aK}^{P_1}\right) \left (C_5-\frac{1}{2}C_7 \right)\right\},
\eeq
where $\xi_u = V_{ub}V_{ud}^*$, $\xi_t = V_{tb}V_{td}^*$.

For $\bar {B}_s^0\to \pi^0 \etap$ decays, the LO decay amplitudes are
\beq
{\cal A}({\bar B}_s^0\to \pi^0\eta) &=&  {\cal A}({\bar B}_s^0\to\pi^0\eta_q)\cos(\phi)
-{\cal A}({\bar B}_s^0\to\pi^0\eta_s)\sin(\phi),\label{eq:pi01}\\
{\cal A}({\bar B}_s^0\to\pi^0\etar)&=&
 {\cal A}({\bar B}_s^0\to\pi^0\eta_q)\sin(\phi)
 +{\cal A}({\bar B}_s^0\to\pi^0\eta_s)\cos(\phi)
,\label{eq:pi02}
\eeq
with
\beq
{\cal A}({\bar B}_s^0\to\pi^0\eta_q)&=& \xi_u^\prime
\left ( f_{B_s}F_{a\eta_n} \; a_2
+ M_{a\eta_n}\; C_2 \right)\non
&&
-\frac{3}{2} \xi_t^\prime \left [ f_{B_s}F_{a\eta_n} \left( a_7+ a_9\right)
+ M_{a\eta_n} \; C_{10} + M_{a\eta_n}^{P_2}\; C_{8} \right], \\
{\cal A}({\bar B}_s^0\to\pi^0\eta_s)&=&
\xi_u^\prime \left (  f_\pi F_{e\eta_s}\; a_2 + M_{e\eta_s}\; C_2\right)\non
&&
-\frac{3}{2} \xi_t^\prime
 \left [ f_\pi F_{e\eta_s} \left (a_9-a_7 \right)
+ M_{a\eta_s} \left ( C_{8} + C_{10} \right ) \right],
\eeq
where $\xi_u^\prime = V_{ub}V_{us}^*$, $\xi_t^\prime = V_{tb}V_{ts}^*$.

For $\bar {B}_s^0\to \eta \eta, \eta \etar, \etar\etar$ decays, the LO decay amplitudes are
\beq
\sqrt{2}{\cal A}({\bar B}_s^0\to\eta\eta)&=&
\cos^2(\phi) {\cal A}({\bar B}_s^0\to\eta_q\eta_q)
-\sin(2\phi){\cal A}({\bar B}_s^0\to\eta_q\eta_s)\non &&
+ \sin^2(\phi) {\cal A}({\bar B}_s^0\to\eta_s\eta_s),  \\
{\cal A}({\bar B}_s^0\to\eta\etar)&=&
\left [ {\cal A}({\bar B}_s^0\to\eta_q\eta_q)
-{\cal A}({\bar B}_s^0\to\eta_s\eta_s)\right ] \cos(\phi)\sin(\phi)
\non &&
+\cos(2\phi){\cal A}({\bar B}_s^0\to\eta_q\eta_s),  \\
\sqrt{2} {\cal A}({\bar B}_s^0\to\etar \etar)&=&
\sin^2(\phi) {\cal A}({\bar B}_s^0\to\eta_q\eta_q)
+\sin(2\phi){\cal A}({\bar B}_s^0\to\eta_q\eta_s)
\non &&
+\cos^2(\phi) {\cal A}({\bar B}_s^0\to\eta_s\eta_s).
\eeq
with
\beq
{\cal A}({\bar B}_s^0\to\eta_q\eta_q)&=&
\xi_u^\prime \; M_{a\eta_n}\; C_2
-\xi_t^\prime \; M_{a\eta_n}\left ( 2C_4 + 2C_6 + \frac{1}{2}C_8 +
\frac{1}{2}C_{10} \right),\\
\sqrt2{\cal A}({\bar B}_s^0\to\eta_q\eta_s)&=&
\xi_u^\prime \left ( f_n F_{e\eta_s}\; a_2
+ M_{e\eta_s}\; C_2 \right)
-\xi_t^\prime \left [ f_n F_{e\eta_s}\left ( 2a_3-2a_5-\frac{1}{2}a_7+\frac{1}{2}a_9 \right)
\right. \non&& \left.
+ M_{a\eta_s}\left (2C_4+2C_6+\frac{1}{2}C_{8}+\frac{1}{2}C_{10} \right) \right ],\\
{\cal A}({\bar B}_s^0\to\eta_s\eta_s)&=&
-2\xi_t^\prime
 \left [ f_s F_{e\eta_s}\left ( a_3+a_4-a_5+\frac{1}{2}a_7-\frac{1}{2}a_9-\frac{1}{2}a_{10}\right)
\right. \non && \left.
+ \left ( f_s F_{e\eta_s}^{P_2} +f_{B_s}F_{a\eta_s}^{P_2}\right)\left ( a_6-\frac{1}{2}a_8 \right)
\right. \non && \left.
+ \left ( M_{e\eta_s} +M_{a\eta_s} \right )
\left (C_3+C_4 +C_6-\frac{1}{2}C_8-\frac{1}{2}C_9-\frac{1}{2}C_{10} \right) \right].
\label{eq:eses}
\eeq

For $\bar{B}^0_{s} \to K^+ \pi^-,K^0 \pi^0$ $K^+ K^-$ and $\bar{K}^0 K^0$ decays,
the LO decay amplitudes are
\beq
{\cal A}(K^+ \pi^-) &=& \xi_u \left (
f_\pi F_{eK}\; a_1 + M_{eK}\; C_1 \right)
- \xi_t \left \{ f_\pi F_{eK} \left ( a_4+a_{10}\right)
+ f_\pi F_{eK}^{P_2} \left ( a_6+a_8 \right)
\right. \non
&& \left.
+ M_{eK}\left ( C_3+C_9 \right)
+ f_{B_s}F_{aK}\left ( a_4-\frac{1}{2}a_{10}\right)
+ f_{B_s} F_{aK}^{P_2}\left ( a_6-\frac{1}{2}a_8\right)
\right. \non && \left.
+ M_{aK}\left (C_3-\frac{1}{2}C_9 \right)
+ M_{aK}^{P_1}\left ( C_5-\frac{1}{2}C_7 \right)\right \},
\eeq
\beq
\sqrt{2}{\cal A}(K^0\pi^0) &=&
\xi_u \left ( f_\pi F_{eK} a_2 + M_{eK} C_2\right)
-\xi_t \left \{
f_\pi F_{eK} \left ( -a_4-\frac{3}{2}a_7+\frac{3}{2}a_9+\frac{1}{2}a_{10} \right)
\right. \non
&& \left.
-\left (  f_\pi F_{ek}^{P_2} + f_{B_s}F_{aK}^{P_2} \right ) \left (a_6-\frac{1}{2}a_8\right)
+ M_{ek}\left ( -C_3 +\frac{3}{2}C_8+\frac{1}{2}C_9+\frac{3}{2}C_{10} \right)
\right. \non
&& \left.
- f_{B_s}F_{ak}\left ( a_4- \frac{1}{2}a_{10}\right)
-M_{ak}\left ( C_3- \frac{1}{2}C_9 \right)
- M_{aK}^{P_1}\left ([C_5-\frac{1}{2}C_7\right) \right\},
\eeq
\beq
{\cal A}(K^+K^-)&=& \xi_u^\prime \left ( f_k F_{ek} a_1 + M_{ek} C_1 + M_{ak} C_2 \right)
\non &&
- \xi_t^\prime \left \{ f_k F_{ek} \left ( a_4+a_{10}\right )
+f_k F_{ek}^{P_2} \left (a_6+a_8 \right)
\right. \non
&& \left.
+ M_{ek} \left (C_3+C_9 \right )
+ M_{ek}^{P_1}\left ( C_5+C_7 \right)
+f_{B_s} F_{ak}^{P_2} \left ( a_6-\frac{1}{2}a_8\right )
\right. \non
&& \left.
+ M_{ak}\left (C_3+C_4 -\frac{1}{2}C_9 -\frac{1}{2}C_{10}\right)
+ M_{ak}^{P_1} \left ( C_5-\frac{1}{2}C_7 \right)
\right. \non
&& \left.
M_{ak}^{P_2}\left (C_6-\frac{1}{2}C_8 \right)
+ \left [ M_{ak} \left ( C_4+C_{10} \right)
+ M_{ak}^{P_2} \left ( C_6+C_8 \right ) \right]_{K^+ \leftrightarrow K^-} \right\},
\eeq
\beq
{\cal A}(\bar K_0 K_0)&=&
- \xi_t^\prime \left \{ f_k F_{ek} \left ( a_4-\frac{1}{2}a_{10} \right)
+ \left( f_k F_{ek}^{P_2} +f_{B_s}F_{ak}^{P_2} \right)
\left ( a_6-\frac{1}{2}a_8 \right)
\right. \non
&& \left.
+ M_{ek}\left ( C_3-\frac{1}{2}C_{9} \right)
+ \left ( M_{ek}^{P_1}+ M_{ak}^{P_1} \right)
\left( C_5- \frac{1}{2}C_{7} \right)
\right. \non
&& \left.
+ M_{ak} \left ( C_3+C_4-\frac{1}{2}C_9 -\frac{1}{2}C_{10} \right)
\right. \non
&& \left.
+ M_{ak} \left ( C_4-\frac{1}{2}C_{10} \right)_{K^0 \leftrightarrow \bar{K}^0}
+ \left [ M_{ak}^{P_2}\left ( C_6-\frac{1}{2}C_8 \right) + [ K^0 \leftrightarrow \bar{K}^0]
\right] \right\}.
\eeq

The $\bar B_s^0\to \pi^+\pi^-$ and $\bar B_s^0\to \pi^0\pi^0$ decays are
pure annihilation processes, and the LO  decay amplitudes can
be written as:
\beq
{\cal A}({\bar B}_s^0\to\pi^+\pi^-)&=&\sqrt2{\cal A}({\bar B}_s^0\to\pi^0\pi^0)\non
&=& \xi_u^\prime M_{a\pi} C_2
-\xi_t^\prime \left [ M_{a\pi}\left ( 2C_4 +  \frac{1}{2}C_{10}\right)
+  M_{a\pi}^{P_2} \left ( 2C_6 + \frac{1}{2}C_8 \right) \right].
\eeq

The individual decay amplitudes, such as $F_{eK}, F_{e\eta_s}, etc.,$
appeared in Eqs.~(\ref{eq:aeta}-\ref{eq:eses}), are obtained by evaluating
the corresponding Feynman diagrams analytically.
For the $B_s \to M_3$ transitions, where the meson $M_3$ absorbed the spectator $\bar{s}$ quark,
the corresponding decay amplitudes can be written as
\beq
F_{eM_3}&=& 8\pi C_F  M_{B_{s}}^{4} \int_0^1 d x_{1} dx_{3}\, \int_{0}^{\infty} b_1 db_1
 b_3 db_3\, \phi_{B_{s}}(x_1,b_1)\non
&& \cdot \left \{ \left [(1+x_3) \phi_3^A(x_3) + r_3 (1-2x_3)
\left (\phi_3^P (x_3)+\phi_3^T (x_3) \right ) \right ]
\right.\non
&& \left.
\quad \cdot \alpha_s(t_e^1)  h_e(x_1,x_3,b_1,b_3)\exp[-S_{ab}(t_e^1)]
\right.\non
&& \left.
+ 2r_3 \phi_3^P (x_3) \cdot \alpha_s(t_e^2)
 h_e(x_3,x_1,b_3,b_1)\exp[-S_{ab}(t_e^2)] \right\} \;,
\label{eq:ab}
\eeq
\beq
F_{eM_3}^{P_{2}} &=&  16\pi C_F   M_{B_{s}}^{4} r_2
\int_0^1 d x_{1} dx_{3}\, \int_{0}^{\infty} b_1 db_1
 b_3 db_3\, \phi_{B_{s}}(x_1,b_1)\non
&& \cdot  \left \{ \left [ \phi_3^A(x_3) + r_3 (2+x_3)\phi_3^P (x_3)-r_3 x_3\phi_3^T (x_3)\right ]
\right.\non
&& \left.
\quad \cdot  \alpha_s(t_e^1)  h_e(x_1,x_3,b_1,b_3)\exp[-S_{ab}(t_e^1)]
\right.\non
&& \left.
+ 2r_3 \phi_3^P (x_3) \alpha_s(t_e^2)
 h_e(x_3,x_1,b_3,b_1)\exp[-S_{ab}(t_e^2)] \right \} \;,
\eeq
\beq
 M_{eM_3}& = & \frac{32} {\sqrt{6}} \pi C_F  M_{B_{s}}^{4}
\int_{0}^{1}\!\!d x_{1}d x_{2}d x_{3}\,\int_{0}^{\infty}\!\! b_1d
b_1 b_2d b_2\, \phi_{B_{s}}(x_1,b_1)
\phi_2^A(x_2)
\non
&& \cdot \left \{\left [(1-x_2) \phi_3^A(x_3)
-r_3x_3 \left (\phi_3^P(x_3)-\phi_3^T(x_3) \right )
\right ]
\right.\non
&& \left.
\quad \cdot   \alpha_s(t_e^3) h_n(x_1,1-x_2,x_3,b_1,b_2)\exp[-S_{cd}(t_e^3)]
\right.\non
&& \left.
- \left [ (x_2+x_3) \phi_3^A (x_3) - r_3 x_3 \left (\phi_3^P(x_3)+\phi_3^T(x_3)\right )\right ]
\right.\non
&& \left.
\quad \cdot \alpha_s(t_e^4)
 h_n(x_1,x_2,x_3,b_1,b_2) \exp[-S_{cd}(t_e^4)] \right \} \; ,
\eeq
\beq
 M_{eM_3}^{P_1}& = & \frac{32} {\sqrt{6}} \pi C_F  M_{B_{s}}^{4}r_2
\int_{0}^{1}d x_{1}d x_{2}d x_{3}\,\int_{0}^{\infty} b_1d b_1 b_2d
b_2\, \phi_{B_{s}}(x_1,b_1) \non
&& \cdot
\left \{ \left [ (1-x_2) \phi_3^A(x_3)\left (\phi_2^P(x_2)+\phi_2^T(x_2)\right )
\right.\right. \non
&&\left. \left.
+ r_3 x_3 \left (\phi_2^P(x_2)-\phi_2^T(x_2) \right )
\left (\phi_3^P(x_3)+\phi_3^T(x_3) \right )
\right.\right. \non
&&\left. \left.
+r_3(1-x_2) \left (\phi_2^P(x_2)+\phi_2^T(x_2) \right )
\left (\phi_3^P(x_3)-\phi_3^T(x_3) \right ) \right ]
\right. \non
&&\left.
\quad \cdot \alpha_s(t_e^3)  h_n(x_1,1-x_2,x_3,b_1,b_2) \exp[-S_{cd}(t_e^3)]
\right. \non
&&\left.
+ \left  [ x_2\phi_3^A (x_3) \left (\phi_2^P(x_2)-\phi_2^T(x_2)\right )
+ r_3 x_2\left (\phi_2^P(x_2)-\phi_2^T(x_2)\right ) \left (\phi_3^P(x_3)-\phi_3^T(x_3)\right )
\right.\right. \non
&&\left.\left.
+ r_3 x_3\left (\phi_2^P(x_2)+\phi_2^T(x_2)\right )
\left (\phi_3^P(x_3)+\phi_3^T(x_3)\right ) \right ]
\right. \non
&&\left.
\quad \cdot \alpha_s(t_e^4) h_n(x_1,x_2,x_3,b_1,b_2)\exp[-S_{cd}(t_e^4)] \right \} \; .
 \eeq
 \beq
 M_{eM_3}^{P_2}& = & \frac{32} {\sqrt{6}} \pi C_F
M_{B_{s}}^{4} \int_{0}^{1}\!d x_{1}d x_{2}d
x_{3}\,\int_{0}^{\infty}\! b_1d b_1 b_2db_2\,\phi_{B_{s}}(x_1,b_1)
\phi_2^A(x_2)\non
&& \cdot \left \{ \left [ (x_2-x_3-1) \phi_3^A(x_3)
+r_3x_3 \left (\phi_3^P(x_3)+\phi_3^T(x_3)\right ) \right ]
\right. \non
&&\left.
\quad \cdot
\alpha_s(t_e^3) h_n(x_1,1-x_2,x_3,b_1,b_2)\exp[-S_{cd}(t_e^3)]
\right. \non
&&\left.
+ \left [x_2\phi_3^A (x_3)+ r_3 x_3 \left (\phi_3^P(x_3)-\phi_3^T(x_3) \right )\right ]
\right. \non
&&\left.
\quad \cdot
 \alpha_s(t_e^4) h_n(x_1,x_2,x_3,b_1,b_2)\exp[-S_{cd}(t_e^4)] \right \} \; ,
\eeq
\beq
F_{aM_3} &=& 8 \pi C_F M_{B_{s}}^{4} \int_{0}^{1} d
x_{2}\,d x_{3}\,\int_{0}^{\infty} b_2d b_2b_3d b_3 \,
  \cdot \left \{ \left [ (x_3-1)\phi_2^A(x_2)\phi_3^{A}(x_3)
\right.\right. \non
&&\left.\left.
  - 4 r_2r_3\phi_2^p(x_2)\phi_3^p(x_3)
  +2r_2r_3x_3\phi_2^p(x_2)\left (\phi_3^P(x_3)-\phi_3^T(x_3)\right )\right ]
\right. \non
&&\left.
\quad \cdot
\alpha_s(t_e^5)  h_a(x_2,1-x_3,b_2,b_3)\exp[-S_{ef}(t_e^5)]
\right. \non
&&\left.
+ \left [x_2 \phi_2^A(x_2) \phi_3^A(x_3)
+ 2 r_2r_3 \left (\phi_2^P(x_2)-\phi_2^T(x_2)\right )\phi_3^P(x_3)
\right.\right. \non
&&\left.\left.
+ 2 r_2 r_3 x_2 \left ( \phi_2^P(x_2)+\phi_2^T(x_2)\right ) \phi_3^P(x_3) \right ]
\right. \non
&&\left.
\quad \cdot   \alpha_s(t_e^6) h_a(1-x_3,x_2,b_3,b_2)\exp[-S_{ef}(t_e^6)] \right \}\;,
\eeq
\beq
F_{aM_3}^{P_2} &=& 16\pi C_F M_{B_{s}}^{4} \int_{0}^{1} d
 x_{2}\,d x_{3}\,\int_{0}^{\infty} b_2d b_2b_3d b_3 \,
\non
&&  \cdot  \left \{ \left [ 2r_2\phi_2^p(x_2)\phi_3^{A}(x_3)
 + (1-x_3)r_3\phi_2^A(x_2) \left (\phi_3^P(x_3)+\phi_3^T(x_3)\right )\right ]
\right. \non
&&\left.
\quad
\cdot   \alpha_s(t_e^5)h_a(x_2,1-x_3,b_2,b_3)\exp[-S_{ef}(t_e^5)]
\right. \non
&&\left.
+\left [ 2 r_3\phi_2^A(x_2) \phi_3^P(x_3)
+ r_2x_2 \left (\phi_2^P(x_2)-\phi_2^T(x_2)\right ) \phi_3^A(x_3)\right ]
\right. \non
&&\left.
\quad \cdot \alpha_s(t_e^6)h_a(1-x_3,x_2,b_3,b_2)\exp[-S_{ef}(t_e^6)] \right \}\;,
\eeq
\beq
M_{aM_3} & = & \frac{32} {\sqrt{6}} \pi C_F  M_{B_{s}}^{4}
\int_{0}^{1}d x_{1}d x_{2}d x_{3}\,\int_{0}^{\infty} b_1d b_1 b_2d
b_2\,\phi_{B_{s}}(x_1,b_1)\non
&&\cdot
\left \{ \left [ -x_2\phi_2^A(x_2)\phi_3^A(x_3)
-4r_2 r_3\phi_2^P(x_2)\phi_3^P(x_3)
\right. \right.\non
&&\left.\left.
+ r_2 r_3(1-x_2) \left ( \phi_2^P(x_2)+\phi_2^T(x_2) \right )
\left(\phi_3^P(x_3)-\phi_3^T(x_3)\right )
\right.\right. \non
&&\left.\left.
+ r_2r_3x_3 \left ( \phi_2^P(x_2)-\phi_2^T(x_2) \right )
\left (\phi_3^P(x_3)+\phi_3^T(x_3)\right )\right ]
\right. \non
&&\left.
\quad \cdot \alpha_s(t_e^7)h_{na}(x_1,x_2,x_3,b_1,b_2)
\exp[-S_{gh}(t_e^7)]
\right. \non
&&\left.
+ \left [ (1-x_3)\phi_2^A(x_2)\phi_3^A(x_3)
+(1-x_3)r_2r_3\left (\phi_2^P(x_2)+\phi_2^T(x_2)\right )
\left (\phi_3^P(x_3)-\phi_3^T(x_3) \right )
\right. \right.\non
&&\left.\left.
+ x_2r_2r_3 \left (\phi_2^P(x_2)-\phi_2^T(x_2) \right )
\left (\phi_3^P(x_3)+\phi_3^T(x_3) \right ) \right ]
\right. \non
&&\left.
\quad \cdot \alpha_s(t_e^8) h'_{na}(x_1,x_2,x_3,b_1,b_2)\exp[-S_{gh}(t_e^8)] \right \} \; ,
\eeq
\beq
 M_{aM_3}^{P_1} & = & \frac{32} {\sqrt{6}} \pi C_F  M_{B_{s}}^{4}
\int_{0}^{1}d x_{1}d x_{2}d x_{3}\,\int_{0}^{\infty} b_1d b_1 b_2d
b_2\,\phi_{B_{s}}(x_1,b_1)\non
&&
\cdot \left \{ \left [ r_2(2-x_2)\left (\phi_2^P(x_2)+\phi_2^T(x_2)\right )\phi_3^A(x_3)
+r_3(1+x_3)\phi_2^A(x_2)\left (\phi_3^P(x_3)-\phi_3^T(x_3)\right ) \right ]
\right. \non
&&\left.
\quad \cdot \alpha_s(t_e^7)\exp[-S_{gh}(t_e^7)] h_{na}(x_1,x_2,x_3,b_1,b_2)
\right. \non
&&\left.
+ \left [ r_2x_2\left (\phi_2^P(x_2)+\phi_2^T(x_2) \right )
\phi_3^A(x_3) -(1-x_3)r_3\phi_2^A(x_2)
\left (\phi_3^P(x_3)-\phi_3^T(x_3) \right )\right ]
\right. \non
&&\left.
\quad \cdot \alpha_s(t_e^8) h'_{na}(x_1,x_2,x_3,b_1,b_2)\exp[-S_{gh}(t_e^8)] \right \} \; ,
\eeq
\beq
 M_{aM_3}^{P_2} & = & \frac{32} {\sqrt{6}} \pi C_F  M_{B_{s}}^{4}
\int_{0}^{1}d x_{1}d x_{2}d x_{3}\,\int_{0}^{\infty} b_1d b_1 b_2d
b_2\,\phi_{B_{s}}(x_1,b_1)\non
&&
\cdot \left \{ \left [ (x_3-1)\phi_2^A(x_2)\phi_3^A(x_3)
-4r_2 r_3\phi_2^P(x_2)\phi_3^P(x_3)
\right. \right.\non
&&\left.\left.
+ r_2 r_3x_3 \left (\phi_2^P(x_2)+\phi_2^T(x_2)\right )
\left (\phi_3^P(x_3)-\phi_3^T(x_3)\right )
\right. \right.\non
&&\left.\left.
+ r_2 r_3 (1-x_2)\left (\phi_2^P(x_2)-\phi_2^T(x_2)\right )
\left (\phi_3^P(x_3)+\phi_3^T(x_3)\right )\right ]
\right. \non
&&\left.
\quad \cdot
\alpha_s(t_e^7) h_{na}(x_1,x_2,x_3,b_1,b_2) \exp[-S_{gh}(t_e^7)]
\right. \non
&&\left.
+ \left [ x_2\phi_2^A(x_2)\phi_3^A(x_3)+x_2r_2r_3\left (\phi_2^P(x_2)+\phi_2^T(x_2)\right )
\left (\phi_3^P(x_3)-\phi_3^T(x_3)\right )
\right. \right.\non
&&\left.\left.
+ r_2 r_3(1-x_3)\left (\phi_2^P(x_2)-\phi_2^T(x_2) \right )
\left ( \phi_3^P(x_3)+\phi_3^T(x_3) \right ) \right ]
\right. \non
&&\left.
\quad \cdot \alpha_s(t_e^8)h'_{na}(x_1,x_2,x_3,b_1,b_2)
\exp[-S_{gh}(t_e^8)] \right \} \; ,
\label{eq:gh}
\eeq
where $C_F=4/3$ is the color-factor,
$r_2 = m_0^{M_2}/M_{B_{s}} $ and $r_3 = m_0^{M_3}/M_{B_{s}}$ with the chiral enhancement
factor $m_0$ for meson $M_2$ and $M_3$.
Here $(F_{eM_3},F_{eM_3}^{P2})$ and $(M_{eM_3},F_{eM_3}^{P1,P2})$  come from the factorizable
emission diagrams ( Fig.1a and 1b) and the non-factorizable hard spectator diagrams
(Fig.1c and 1d), respectively; and
$(F_{aM_3},F_{aM_3}^{P2})$ and $(M_{aM_3},F_{aM_3}^{P1,P2})$  are obtained by
evaluating the factorizable annihilation diagrams ( Fig.1e and 1f) and the non-factorizable
annihilation diagrams (Fig.1g and 1h), respectively.
The explicit expressions of the hard energy scale $(t_e^1,t_e^2,\cdots,t_e^8)$, the hard
functions $(h_e, h_n, h_a, h_{na}, h^\prime_{na})$, the Sudakov factors
$(S_{ab}(t_e), S_{cd}(t_e),S_{ef}(t_e),S_{gh}(t_e))$ can be found in Appendix A.

\section{Next-to-leading order contributions}\label{sec:nlo}

At next-to-leading order, firstly, the NLO Wilson coefficients $C_i(\mw)$, the NLO
RG evolution matrix $U(t,m,\alpha)$ \cite{buras96}, and the $\alpha_s(t)$ at
two-loop level will be employed.

\begin{figure}[tb]
\vspace{-5cm} \centerline{\epsfxsize=18 cm \epsffile{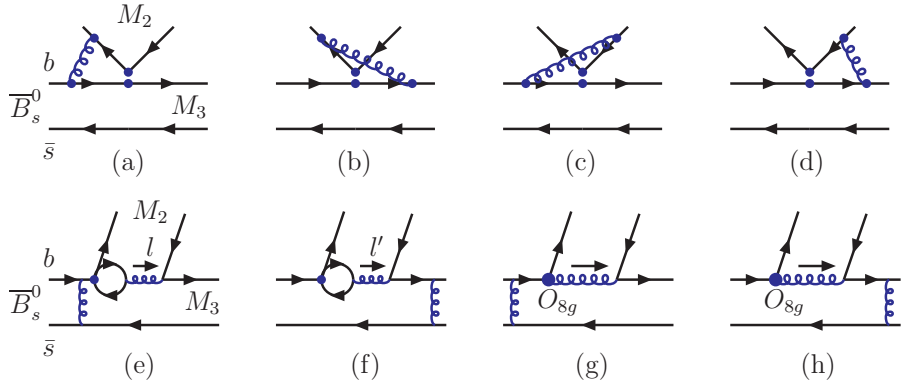}}
\vspace{-15cm}
\caption{Feynman diagrams for NLO contributions:  the vertex corrections (a-d);
the quark-loop (e-f) and the chromo-magnetic penguin contributions (g-h).}
\label{fig:fig2}
\end{figure}

\begin{figure}[tb]
\vspace{-5cm}
\centerline{\epsfxsize=18 cm \epsffile{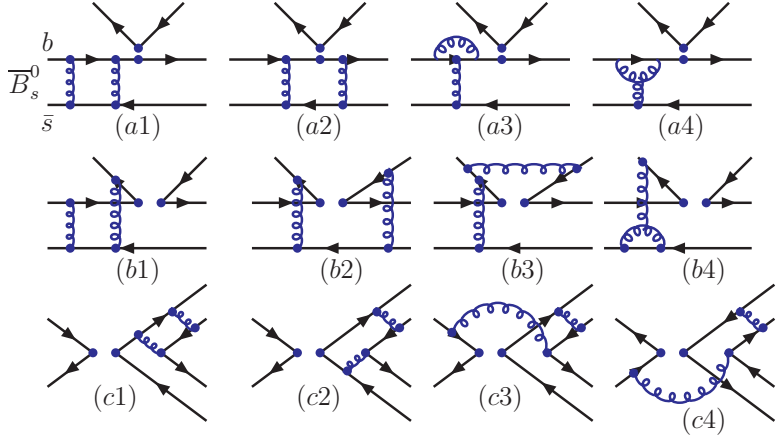}}
\vspace{-14cm}
\caption{The typical Feynman diagrams, which provide NLO contributions to the
factorizable emission amplitudes (a1-a4),  the hard-spectator amplitude (b1-b4),
and the annihilation amplitudes (c1-c4)}
\label{fig:fig3}
\end{figure}

Secondly, the NLO hard kernel $H^{(1)}(\alpha_s^2)$ should be included.
All the Feynman diagrams,  in other words, which lead to the decay amplitudes
proportional to $\alpha^2_s(t)$, should be considered.
Such Feynman diagrams can be grouped into following classes:
\begin{itemize}
\item[]{I:}
The vertex corrections, as illustrated in Figs.~\ref{fig:fig2}a-\ref{fig:fig2}d,
the same set as in the QCDF approach.

\item[]{II:}
The NLO contributions from quark-loops, as illustrated  in Figs.~\ref{fig:fig2}e-\ref{fig:fig2}f.

\item[]{III:}
The NLO contributions from chromo-magnetic penguins, i.e. the operator $O_{8g}$,
as illustrated  in Figs.~\ref{fig:fig2}g-\ref{fig:fig2}h.
There are totally nine relevant Feynman diagrams
as given in Ref.~\cite{o8g2003}, if the Feynman diagrams involving
three-gluon vertex are also included.
We here show the first two only, and they provide
the dominant NLO contributions, according to Ref.~\cite{o8g2003}.

\item[]{IV:}
The NLO corrections to the LO emission diagrams(1a,1b), the LO hard-spectator(1c,1d)
and the LO annihilation diagrams (1e-1h), as illustrated in Fig.~\ref{fig:fig3}.
One can draw more than one hundred such Feynman diagrams in total,
but we here show representative ones only.

\end{itemize}

At present, the calculations for the vertex corrections, the quark-loops and
chromo-magnetic penguins have been available and will be considered here.
We expect that they are the major part of the full
NLO contributions in pQCD approach \cite{nlo05}.
For the Feynman diagrams as shown in Fig.~3, however,
the analytical calculations have not been completed yet.
Of course, these Feynman diagrams should be calculated as soon as possible, in order to
improve the reliability of the pQCD predictions.

\subsection{Vertex corrections, quark-loops and chromo-magnetic penguins}\label{sec:vc2}

The vertex corrections to the factorizable emission diagrams, as illustrated by
Figs.~2a-2d, have been calculated years ago in the QCD factorization
approach\cite{bbns99,npb675}.
For $B_s \to PP$ decays, the vertex corrections can be calculated without
considering the transverse momentum effects of the quark at the
end-point \cite{nlo05}, one can use the vertex corrections as given in
Ref.~\cite{npb675} directly.
The vertex corrections can then be absorbed into the re-definition of the Wilson coefficients
$a_i(\mu)$ by adding a vertex-function $V_i(M)$ to them
\cite{bbns99,npb675}
\beq
a_i(\mu)&\to & a_i(\mu) +\frac{\alpha_s(\mu)}{4\pi}C_F\frac{C_i(\mu)}{3} V_i(M),\ \ for
\ \ i=1,2; \non
a_j(\mu)&\to & a_j(\mu)+\frac{\alpha_s(\mu)}{4\pi}C_F\frac{C_{j+1}(\mu)}{3}
V_j(M), \ \  for \ \ j=3,5,7,9, \non
a_j(\mu)&\to & a_j(\mu)+\frac{\alpha_s(\mu)}{4\pi}C_F\frac{C_{j-1}(\mu)}{3}
V_j(M), \ \  for \ \ j=4,6,8,10,
\label{eq:aimu-2}
\eeq
where M is the meson emitted from the weak vertex. When $M$ is a
pseudo-scalar meson, the vertex functions $V_{i}(M)$ are given ( in the NDR
scheme) in Refs.~\cite{npb675,nlo05}:
\beq
V_i(M)&=&\left\{ \begin{array}{cc}
12\ln\frac{m_b}{\mu}-18+\frac{2\sqrt{6}}{f_M}\int_{0}^{1}dx\phi_M^A(x)g(x),
& {\rm for}\quad i= 1-4,9,10,\\
-12\ln\frac{m_b}{\mu}+6-\frac{2\sqrt{6}}{f_M}\int_{0}^{1}dx\phi_M^A(x)g(1-x),
&{\rm for}\quad i= 5,7,\\
-6+\frac{2\sqrt{6}}{f_M}\int_{0}^{1}dx\phi_M^P(x)h(x),
&{\rm for} \quad i= 6,8,\\
\end{array}\right.
\label{eq:vim}
\eeq
where $f_M$ is the decay constant of the meson M; $\phi_M^A(x)$ and $\phi_M^P(x)$ are
the twist-2 and twist-3 distribution amplitude of the meson M, respectively.
The hard-scattering
functions $g(x)$ and $h(x)$ in Eq.~(\ref{eq:vim}) are:
\beq
g(x)&=& 3\left(\frac{1-2x}{1-x}\ln x-i\pi\right)\non
&& +\left[2Li_2(x)-\ln^2x+\frac{2\ln
x}{1-x}-\left(3+2i\pi\right)\ln x -(x\leftrightarrow 1-x)\right],\\
h(x)&=&2Li_2(x)-\ln^2x-(1+2i\pi)\ln x-(x\leftrightarrow 1-x),
\eeq
where ${\rm Li}_2(x)$ is the dilogarithm function. As shown in Ref.~\cite{nlo05},
the $\mu$-dependence of the Wilson coefficients $a_i(\mu)$ will be improved generally
by the inclusion of the vertex corrections.


The contribution from the so-called ``quark-loops" is a kind of penguin correction
with the four quark operators insertion, as illustrated by Fig.~\ref{fig:fig2}e and
\ref{fig:fig2}f.
We here include quark-loop amplitude from the
operators $O_{1,2}$ and $O_{3-6}$ only. The quark loops from $O_{7-10}$
will be neglected due to their smallness.

For the $b\to s$ transition, the effective Hamiltonian $H_{eff}^{ql}$ which describes
the contributions from the quark loops can be written as \cite{nlo05}
\beq
H_{eff}^{(ql)}&=&-\sum\limits_{q=u,c,t}\sum\limits_{q{\prime}}\frac{G_F}{\sqrt{2}}
V_{qb}V_{qs}^{*}\frac{\alpha_s(\mu)}{2\pi}C^{(q)}(\mu,l^2)\left(\bar{s}\gamma_\rho
\left(1-\gamma_5\right)T^ab\right)\left(\bar{q}^{\prime}\gamma^\rho
T^a q^{\prime}\right),
\eeq
where $l^2$ is  the invariant mass of the gluon, as illustrated by
Fig.\ref{fig:fig2}e. The functions $C^{(q)}(\mu,l^2)$ are given by
\beq
C^{(q)}(\mu,l^2)&=&\left[G^{(q)}(\mu,l^2)-\frac{2}{3}\right]C_2(\mu)
\label{eq:qlc}
\eeq
for $q=u,c$ and
\beq
C^{(t)}(\mu,l^2)&=&\left[G^{(s)}(\mu,l^2)-\frac{2}{3}\right]
C_3(\mu)+\sum\limits_{q^\prime=u,d,s,c}G^{(q^\prime)}(\mu,l^2)
\left[C_4(\mu)+C_6(\mu)\right]\label{eq:eh}
\eeq
for $q=t$.
The function $G^{(q)}(\mu,l^2)$ for the loop of the $q(q=u,d,s,c)$ quark is
given by \cite{nlo05}
\beq
G^{(q)}(\mu,l^2)&=&-4\int_{0}^{1}dx x(1-x)\ln\frac{m_q^2-x(1-x)l^2}{\mu^2}
\eeq
$m_q$ is the possible quark mass. The explicit  expressions of the function
$G^{(q)}(\mu,l^2)$ after the integration can be found, for example, in
Ref.~\cite{nlo05}.

The magnetic penguin is another kind penguin correction induced by the insertion of
the operator $O_{8g}$, as illustrated by Fig.2g and 2h. The corresponding weak
effective Hamiltonian contains the $b\to s g$ transition can be written as
\beq
H_{eff}^{mp} &=&-\frac{G_F}{\sqrt{2}} V_{tb}V_{ts}^*\; C_{8g}^{eff} O_{8g},
\eeq
with the magnetic penguin operator,
\beq
O_{8g}&=&\frac{g_s}{8\pi^2}m_b\; \bar{d}_i \; \sigma^{\mu\nu}\; (1+\gamma_5)\;
 T^a_{ij}\; G^a_{\mu\nu}\;  b_j,
\label{eq:o8g}
\eeq
where $i,j$ being the color indices of quarks. The corresponding effective Wilson
coefficient $C_{8g}^{eff}= C_{8g} + C_5$ \cite{nlo05}.

It is worth noting that there are totally nine Feynman diagrams involving
the $O_{8g}$ operator as shown in Ref.~\cite{o8g2003}, if the Feynman diagrams involving
three-gluon vertex are also included. We here show the first two only, say Fig.2g and 2h,
and they provide the dominant NLO contributions from the $O_{8g}$ operator,
according to Ref.~\cite{o8g2003}.


\subsection{NLO decay amplitudes}\label{sec:nlo2}

Now we are ready to calculate the decay amplitude corresponding to
the quark-loops and magnetic penguins. By analytical evaluations, we
find two kinds of topological decay amplitudes $\calm_{M_2M_3}^{q}$
and $\calm_{M_2M_3}^{g}$, respectively.
For $B \to \eta_n K^0$ decay, we find
\beq
\calm^{(q)}_{\eta_n K^0}&=& -\frac{4}{\sqrt{3}} m_{B_s}^4C_F^2\;  \int_0^1
dx_1dx_2dx_3 \int_0^\infty b_1db_1b_3db_3 \,\phi_{B_s}(x_1)
\left \{ \left [ (1+x_3)\phi_{\eta_n}^A(x_2) \phi_K^A(x_3)
\right. \right. \non
&& \left. \left.
+ 2 r_{\eta_n}\phi_{\eta_n}^P(x_2) \phi_K^A(x_3)+
r_K(1-2x_3)\phi_{\eta_n}(x_2) \left [ \phi_K^P(x_3)+\phi_K^T(x_3) \right]
\right.\right.\non &&
\left.\left.
+ 2r_{\eta_n} r_K \phi_{\eta_n}^P(x_2)
\left [ (2+x_3)\phi_K^P(x_3) - x_3\phi_K^T(x_3) \right ] \right ]
\right.\non &&
\left.
\cdot \alpha_s^2(t_e^9) h_e(x_1,x_3,b_1,b_3) \exp[-S_{ab}(t_e^9)]C^{(q)}(t_e^9,l^2)
\right.
\non && \left.
+ \left [2r_K\phi_{\eta_n}^A(x_2)\phi_K^P(x_3) +
4r_{\eta_n}r_K\phi_{\eta_n}^P(x_2)\phi_K^P(x_3)\right ]
\right. \non &&
\left.
\cdot \alpha_s^2(t_e^{10}) h_e(x_3,x_1,b_3,b_1) \exp[-S_{ab}(t_e^{10})]
C^{(q)}(t_e^{10},l'^2)\right \},
\label{eq:mqketan}
\eeq
\beq
\calm^{(g)}_{\eta_n K^0} &=&
\frac{8 m_{B_s}^6 C_F^2 }{\sqrt{3}} \int_0^1
dx_1dx_2dx_3 \int_0^\infty b_1db_1b_2db_2b_3db_3\, \phi_{B_s}(x_1)
\left \{ \left [-(1-x_3) [ \phi_K^A(x_3)
\right. \right.
\non && \left. \left.
+ r_K(3\phi_K^P(x_3) +\phi_K^T(x_3)) + r_Kx_3(\phi_K^P(x_3)-\phi_K^T(x_3))
 ] \phi_{\eta_n}^A(x_2) - r_{\eta_n}x_2(1+x_3)
\right.\right.
\non && \left.\left.
\times (3\phi_{\eta_n}^P(x_2) -\phi_{\eta_n}^T(x_2))\phi_K^A(x_3)
-r_{\eta_n}r_K(1-x_3)(3\phi_{\eta_n}^P(x_2)+\phi_{\eta_n}^T(x_2))(\phi_K^P(x_3)
\right.\right.
\non && \left. \left.
- \phi_K^T(x_3)) - r_{\eta_n}r_K x_2 (1-2x_3)(3\phi_{\eta_n}^P(x_2) -
\phi_{\eta_n}^T(x_2))(\phi_K^P(x_3) + \phi_K^T(x_3)) \right ]
\right.
\non && \left.
\cdot \alpha_s^2(t_e^{9}) h_g(x_1,x_2,x_3,b_1,b_2,b_3) \exp[-S_{cd}(t_e^{9})]
C_{8g}^{eff}(t_e^{9})
\right.
\non && \left.
 - \left [ 4r_K\phi_{\eta_n}^A(x_2)\phi_K^P(x_3)
 + 2r_{\eta_n}r_K x_2(3\phi_{\eta_n}^P(x_2)- \phi_{\eta_n}^T(x_2))
 \phi_K^P(x_3) \right]
\right.
\non && \left.
\cdot \alpha_s^2(t_e^{10}) h'_g(x_1,x_2,x_3,b_1,b_2,b_3)
\exp[-S_{cd}(t_e^{10})] C_{8g}^{eff}(t_e^{10})\right \}.
\label{eq:mgketan}
\eeq
\beq
{\cal M}^{(q)}_{K^0 \eta_s}&=&
-\frac{8m_{B_s}^4 C_F^2}{\sqrt{6}} \int_0^1 dx_1dx_2dx_3
\int_0^\infty b_1db_1b_3db_3 \,\phi_{B_s}(x_1)
\left \{\left [ (1+x_3)\phi_{K}^A(x_2) \phi_{\eta_s}^A(x_3)
\right.\right.
\non && \left.\left.
+ 2 r_{K}\phi_K^P(x_2) \phi_{\eta_s}^A(x_3)+
r_{\eta_s}(1-2x_3)\phi_K(x_2) \left[ \phi_{\eta_s}^P(x_3)
+ \phi_{\eta_s}^T(x_3) \right]
\right.\right.
\non && \left.\left.
+ 2r_K r_{\eta_s} \phi_K^P(x_2)
\left[ (2+x_3)\phi_{\eta_s}^P(x_3) -x_3\phi_{\eta_s}^T(x_3) \right ] \right ]
\right.
\non && \left.
\cdot  \alpha_s^2(t_e^9) h_e(x_1,x_3,b_1,b_3)
\exp[-S_{ab}(t_e^9)] C^{(q)}(t_e^9,l^2)
\right.
\non && \left.
+ \left [2r_{\eta_s}\phi_K^A(x_2)\phi_{\eta_s}^P(x_3) + 4r_K
r_{\eta_s}\phi_K^P(x_2)\phi_{\eta_s}^P(x_3) \right]
\right.
\non && \left.
\cdot  \alpha_s^2(t_e^{10})h_e(x_3,x_1,b_3,b_1) \exp[-S_{ab}(t_e^{10})]
C^{(q)}(t_e^{10},l'^2)\right \},
\label{eq:mqketas}
\eeq
\beq
{\cal M}^{(g)}_{K^0 \eta_s} &=& 16m_{B_s}^6\frac{{C_F}^2}{\sqrt{2N_c}} \int_0^1 dx_1dx_2dx_3
\int_0^\infty b_1db_1b_2db_2b_3db_3\, \phi_{B_s}(x_1)
\non
&& \cdot \left \{\left [-(1-x_3) \left [ 2\phi_{\eta_s}^A(x_3)
+ r_{\eta_s} \left [ 3\phi_{\eta_s}^P(x_3) + \phi_{\eta_s}^T(x_3) \right]
\right.\right.\right.
\non && \left.\left.\left.
+ r_{\eta_s}x_3(\phi_{\eta_s}^P(x_3)
-\phi_{\eta_s}^T(x_3)) \right] \phi_K^A(x_2)
\right.\right.
\non && \left.\left.
 - r_K x_2(1+x_3)\left [3\phi_K^P(x_2)
 -\phi_K^T(x_2) \right] \phi_{\eta_s}^A(x_3)
\right.\right.
\non && \left.\left.
 -r_K r_{\eta_s}(1-x_3)(3\phi_K^P(x_2)
+ \phi_K^T(x_2))(\phi_{\eta_s}^P(x_3)
\right.\right.
\non && \left.\left.
- \phi_{\eta_s}^T(x_3)) - r_Kr_{\eta_s} x_2 (1-2x_3)(3\phi_K^P(x_2)
- \phi_K^T(x_2))(\phi_{\eta_s}^P(x_3)
+ \phi_{\eta_s}^T(x_3) ) \right ]
\right.
\non && \left.
\cdot \alpha_s^2(t_e^{9}) C_{8g}^{eff}(t_e^{9}) h_g(x_1,x_2,x_3,b_1,b_2,b_3)
\exp[-S_{cd}(t_e^{9})]
\right.
\non && \left.
- \left [ 4r_{\eta_s}\phi_K^A(x_2)\phi_{\eta_s}^P(x_3)
+ 2r_K r_{\eta_s} x_2(3\phi_K^P(x_2)
- \phi_K^T(x_2))\phi_{\eta_s}^P(x_3) \right ]
\right.
\non && \left.
\cdot \alpha_s^2(t_e^{10}) h'_g(x_1,x_2,x_3,b_1,b_2,b_3) \exp[-S_{cd}(t_e^{10})]
C_{8g}^{eff}(t_e^{10})\right\},
\label{eq:mgketas}
\eeq
\beq
{\cal M}^{(q)}_{\eta_s \eta_s}&=& -16m_{B_s}^4\frac{{C_F}^2}{\sqrt{2N_c}} \int_0^1 dx_1dx_2dx_3
\int_0^\infty b_1db_1b_3db_3 \,\phi_{B_s}(x_1)
\left \{ \left [(1+x_3)\phi_{\eta_s}^A(x_2) \phi_{\eta_s}^A(x_3)
\right.\right.
\non && \left.\left.
+ 2 r_{\eta_s}\phi_{\eta_s}^P(x_2) \phi_{\eta_s}^A(x_3)+
r_{\eta_s}(1-2x_3)\phi_{\eta_s}(x_2)(\phi_{\eta_s}^P(x_3)
+ \phi_{\eta_s}^T(x_3))
\right.\right.
\non && \left.\left.
+ 2r_{\eta_s} r_{\eta_s} \phi_{\eta_s}^P(x_2) ((2+x_3)\phi_{\eta_s}^P(x_3)
-x_3\phi_{\eta_s}^T(x_3) ) \right]
\right.
\non && \left.
\cdot \alpha_s^2(te^9) h_e(x_1,x_3,b_1,b_3) \exp[-S_{ab}(t_e^9)] C^{(q)}(t_e^9,l^2)
\right.
\non && \left.
+ \left [ 2r_{\eta_s}\phi_{\eta_s}^A(x_2)\phi_{\eta_s}^P(x_3) + 4r_{\eta_s}
r_{\eta_s}\phi_{\eta_S}^P(x_2)\phi_{\eta_s}^P(x_3)
\right ]
\right.
\non && \left.
\cdot \alpha_s^2(t_e^{10}) h_e(x_3,x_1,b_3,b_1)
\exp[-S_{ab}(t_e^{10})] C^{(q)}(t_e^{10},l'^2)\right\},
\eeq
\beq
{\cal M}^{(g)}_{\eta_s \eta_s} &=& -32m_{B_s}^6\frac{{C_F}^2}{\sqrt{2N_c}} \int_0^1 dx_1dx_2dx_3
\int_0^\infty b_1db_1b_2db_2b_3db_3\, \phi_{B_s}(x_1)
\left \{ \left [(1-x_3) \left [ 2\phi_{\eta_s}^A(x_3)
\right.\right.\right.
\non && \left.\left.\left.
+ r_{\eta_s}(3\phi_{\eta_s}^P(x_3)
+\phi_{\eta_s}^T(x_3) )
+ r_{\eta_s}x_3(\phi_{\eta_s}^P(x_3)
-\phi_{\eta_s}^T(x_3))\right] \phi_{\eta_s}^A(x_2)
\right.\right.
\non && \left.\left.
- r_{\eta_s} x_2(1+x_3) (3\phi_{\eta_s}^P(x_2)
-\phi_{\eta_s}^T(x_2))\phi_{\eta_s}^A(x_3)
\right.\right.
\non && \left.\left.
-r_{\eta_s}r_{\eta_s}(1-x_3)(3\phi_{\eta_s}^P(x_2)
+\phi_{\eta_s}^T(x_2))(\phi_{\eta_s}^P(x_3) -\phi_{\eta_s}^T(x_3))
\right.\right.
\non && \left.\left.
- r_{\eta_s}r_{\eta_s} x_2 (1-2x_3)(3\phi_{\eta_s}^P(x_2)
- \phi_{\eta_s}^T(x_2))(\phi_{\eta_s}^P(x_3)
+ \phi_{\eta_s}^T(x_3)) \right ]
\right.
\non &&\left.
\cdot \alpha_s^2(t_e^{9}) h_g(x_1,x_2,x_3,b_1,b_2,b_3)
\exp[-S_{cd}(t_e^{9})]  C_{8g}^{eff}(t_e^{9})
\right.
\non && \left.
+ \left [4r_{\eta_s}\phi_{\eta_s}^A(x_2)\phi_{\eta_s}^P(x_3) + 2r_{\eta_s}
r_{\eta_s} x_2(3\phi_{\eta_s}^P(x_2)
- \phi_{\eta_s}^T(x_2))\phi_{\eta_s}^P(x_3) \right]
\right.
\non && \left.
\cdot \alpha_s^2(t_e^{10}) h'_g(x_1,x_2,x_3,b_1,b_2,b_3)
\exp[-S_{cd}(t_e^{10})] C_{8g}^{eff}(t_e^{10})\right\}.
\eeq

It is easy to see that the decay modes $B_s^0 \to \pi^0 \eta_n, \pi^0 \eta_s, \eta_n
\eta_n $ and  $\eta_n \eta_s$,
do not receive the NLO contributions from the quark-loop and the magnetic-penguin
diagrams.

For other $B_s \to PP$ decays involving no $\etap$ mesons, the corresponding
quark-loop and magnetic penguin amplitudes are of the form
\beq
{\cal M}^{(q)}_{\pi^- K^+}&=&
-8m_{B_s}^4\frac{{C_F}^2}{\sqrt{2N_c}} \int_0^1 dx_1dx_2dx_3
\int_0^\infty b_1db_1b_3db_3 \,\phi_{B_s}(x_1)
\left \{ \left [ (1+x_3)\phi_{\pi}^A(x_2) \phi_K^A(x_3)
\right.\right.
\non && \left.\left.
+ 2r_{\pi}\phi_{\pi}^P(x_2) \phi_K^A(x_3)+
r_K(1-2x_3)\phi_{\pi}(x_2)(\phi_K^P(x_3)+\phi_K^T(x_3))
\right.\right.
\non && \left.\left.
+ 2r_{\pi} r_K \phi_{\pi}^P(x_2) ((2+x_3)\phi_K^P(x_3)
-x_3\phi_K^T(x_3)) \right ]
\right.
\non && \left.
\cdot \alpha_s^2(t_e^9) h_e(x_1,x_3,b_1,b_3)
\exp[-S_{ab}(t_e^9)] C^{(q)}(t_e^9,l^2)
\right.
\non && \left.
+ \left [ 2r_K\phi_{\pi}^A(x_2)\phi_K^P(x_3) +
4r_{\pi}r_K\phi_{\pi}^P(x_2)\phi_K^P(x_3) \right ]
\right.
\non && \left.
\cdot \alpha_s^2(t_e^{10}) h_e(x_3,x_1,b_3,b_1)
 \exp[-S_{ab}(t_e^{10})] C^{(q)}(t_e^{10},l'^2)\right \},
\label{eq:mqkpi}
\eeq
\beq
{\cal M}^{(g)}_{\pi^- K^+} &=&-16m_{B_s}^6\frac{{C_F}^2}{\sqrt{2N_c}} \int_0^1 dx_1dx_2dx_3
\int_0^\infty b_1db_1b_2db_2b_3db_3\, \phi_{B_s}(x_1)
\non &&
\cdot \left \{ \left [ (1-x_3) \left [ 2\phi_K^A(x_3) +
r_K(3\phi_K^P(x_3) +\phi_K^T(x_3) )
\right.\right.\right.
\non && \left.\left.\left.
+ r_Kx_3(\phi_K^P(x_3)-\phi_K^T(x_3)) \right ]
\phi_{\pi}^A(x_2)
\right.\right.
\non && \left.\left.
- r_{\pi}x_2(1+x_3) (3\phi_{\pi}^P(x_2) -\phi_{\pi}^T(x_2))\phi_K^A(x_3)
\right.\right.
\non && \left.\left.
-r_{\pi}r_K(1-x_3)(3\phi_{\pi}^P(x_2)
+ \phi_{\pi}^T(x_2))(\phi_K^P(x_3) -\phi_K^T(x_3))
\right.\right.
\non && \left.\left.
- r_{\pi}r_K x_2 (1-2x_3)(3\phi_{\pi}^P(x_2) -
\phi_{\pi}^T(x_2))(\phi_K^P(x_3) + \phi_K^T(x_3)) \right ]
\right.
\non &&\left.
\cdot \alpha_s^2(t_e^{9})
h_g(x_1,x_2,x_3,b_1,b_2,b_3) \exp[-S_{cd}(t_e^{9})] C_{8g}^{eff}(te^{9})
\right.
\non &&\left.
+ \left [ 4r_K\phi_{\pi}^A(x_2)\phi_K^P(x_3)
+ 2r_{\pi}r_K x_2(3\phi_{\pi}^P(x_2)
- \phi_{\pi}^T(x_2))\phi_K^P(x_3)\right ]
\right.
\non &&\left.
\alpha_s^2(t_e^{10}) h'_g(x_1,x_2,x_3,b_1,b_2,b_3)
\exp[-S_{cd}(t_e^{10})] C_{8g}^{eff}(t_e^{10})\right\},
\label{eq:mgkpi}
\eeq
\beq
{\cal M}^{(q)}_{K^0 \pi^0}&=& \frac{1}{\sqrt {2}}{\cal M}^{(q)}_{\pi^- K^+},\\
{\cal M}^{(g)}_{K^0 \pi^0}&=& \frac{1}{\sqrt {2}}{\cal M}^{(g)}_{\pi^- K^+},
\eeq
\beq
 {\cal M}^{(q)}_{K^+ K^-}&=&
-8m_{B_s}^4\frac{{C_F}^2}{\sqrt{2N_c}} \int_0^1 dx_1dx_2dx_3
\int_0^\infty b_1db_1b_3db_3 \,\phi_{B_s}(x_1)
\non &&
\cdot \left \{\left [(1+x_3)\phi_{k}^A(x_2) \phi_K^A(x_3)
+2r_{k}\phi_{k}^P(x_2) \phi_K^A(x_3)
\right.\right.
\non && \left.\left.
+ r_K(1-2x_3)\phi_{k}(x_2)(\phi_K^P(x_3)+\phi_K^T(x_3))
\right.\right.
\non && \left.\left.
+ 2r_{k} r_K \phi_{k}^P(x_2)((2+x_3)\phi_K^P(x_3)-x_3\phi_K^T(x_3))\right ]
\right.
\non &&\left.
\cdot \alpha_s^2(t_e^9) h_e(x_1,x_3,b_1,b_3)
\exp[-S_{ab}(t_e^9)] C^{(q)}(t_e^9,l^2)
\right.
\non && \left.
+ \left [2r_K\phi_k^A(x_2)\phi_K^P(x_3) +
4r_kr_K\phi_k^P(x_2)\phi_K^P(x_3)\right ]
\right.
\non && \left.
\cdot \alpha_s^2(t_e^{10}) h_e(x_3,x_1,b_3,b_1)
\exp[-S_{ab}(t_e^{10})] C^{(q)}(t_e^{10},l'^2)\right\},
 \label{eq:mqkk}
\eeq
\beq {\cal M}^{(g)}_{K^+ K^-} &=&
- 16m_{B_s}^6\frac{{C_F}^2}{\sqrt{2N_c}} \int_0^1 dx_1dx_2dx_3
\int_0^\infty b_1db_1b_2db_2b_3db_3\, \phi_{B_s}(x_1)
\non
&&
\cdot \left \{ \left [(1-x_3) \left [2\phi_K^A(x_3) + r_K(3\phi_K^P(x_3) +\phi_K^T(x_3))
\right.\right.\right.
\non && \left.\left.\left.
+r_Kx_3(\phi_K^P(x_3)-\phi_K^T(x_3)) \right] \phi_K^A(x_2)
\right.\right.
\non && \left.\left.
- r_Kx_2(1+x_3)(3\phi_K^P(x_2) -\phi_K^T(x_2))\phi_K^A(x_3)
\right.\right.
\non && \left.\left.
-r_K^2(1-x_3)(3\phi_K^P(x_2) + \phi_K^T(x_2))(\phi_K^P(x_3) -\phi_K^T(x_3))
\right.\right.
\non && \left.\left.
- r_K^2 x_2 (1-2x_3)(3\phi_K^P(x_2) - \phi_K^T(x_2))(\phi_K^P(x_3) +
\phi_K^T(x_3)) \right ]
\right.
\non &&\left.
\cdot
\alpha_s^2(t_e^{9}) h_g(x_1,x_2,x_3,b_1,b_2,b_3) \exp[-S_{cd}(t_e^{9})]
C_{8g}^{eff}(t_e^{9})
\right.
\non && \left.
+ \left [4r_K\phi_K^A(x_2)\phi_K^P(x_3) + 2r_Kr_K x_2(3\phi_K^P(x_2)
- \phi_K^T(x_2))\phi_K^P(x_3) \right ]
\right.
\non &&\left.
\cdot \alpha_s^2(t_e^{10}) h'_g(x_1,x_2,x_3,b_1,b_2,b_3)
\exp[-S_{cd}(t_e^{10})] C_{8g}^{eff}(t_e^{10})\right\},
\label{eq:mgkk}
\eeq
\beq
{\cal M}^{(q)}_{\bar K^0 k^0} &=& {\cal M}^{(q)}_{k^+ k^-}, \\
{\cal M}^{(g)}_{\bar K^0 k^0} &=& {\cal M}^{(g)}_{k^+ k^-}.
 \eeq
The functions $h_e, h_g, h_g^\prime$, the hard scales $t_e^9$ and $t_e^{10}$, the
Sudakov factors $S_{ab}(t)$ and $S_{cd}(t)$, appeared  in
Eqs.(\ref{eq:mqketan})-(\ref{eq:mgkk}), will be given in Appendix A.

For the pure annihilation-type decays $B_s^0 \to \pi^+ \pi^-$ and $B_s^0 \to \pi^0 \pi^0$,
they do not receive the NLO contributions from the vertex corrections, the quark loops  and
the magnetic-penguins.
For $B_s^0 \to \pi^0 \eta$ and  $B_s^0 \to \pi^0 \etar$
decays, the quark-loops and magnetic penguins also do not contribute.
For other decays, the NLO contributions will be included in a simple way:
\beq
{\cal A}_{\eta_n K^0} &\to& {\cal A}_{\eta_n K^0}
+\sum_{q=u,c,t} \xi_q{\cal M}_{\eta_n K^0}^{(q)} +\xi_t{\cal M}_{\eta_n K^0}^{(g)}, \non
{\cal A}_{K^0 \eta_s} &\to& {\cal A}_{K^0 \eta_s}
+\sum_{q=u,c,t} \xi_q {\cal M}_{K^0 \eta_s}^{(q)} + \xi_t{\cal M}_{K^0 \eta_s }^{(g)},  \non
{\cal A}_{\eta_s \eta_s} &\to& {\cal A}_{\eta_s \eta_s}
+\sum_{q=u,c,t} \xi^{\prime}_q {\cal M}_{\eta_s \eta_s}^{(q)}
+\xi^{\prime}_t{\cal M}_{\eta_s \eta_s}^{(g)},\non
{\cal A}_{\pi^- K^+} &\to & {\cal A}_{\pi^- K^+} +
\sum_{q=u,c,t} \xi_q{\cal M}_{\pi^- K^+}^{(q)} + \xi_t{\cal M}_{\pi^- K^+}^{(g)}, \non
{\cal A}_{\pi^0 K^0} &\to & {\cal A}_{\pi^0 K^0}
+\sum_{q=u,c,t} \xi_q{\cal M}_{\pi^0 K^0}^{(q)}+ \xi_t{\cal M}_{\pi^0 K^0}^{(g)},\non
{\cal A}_{K^- K^+} &\to & {\cal A}_{K^- K^+}
+ \sum_{q=u,c,t} \xi^{\prime}_q{\cal M}_{K^- K^+}^{(q)}
+ \xi^{\prime}_t{\cal M}_{K^- K^+}^{(g)},\non
{\cal A}_{\bar K^0 K^0} &\to & {\cal A}_{\bar K^0 K^0}
+ \sum_{q=u,c,t} \xi^{\prime}_q{\cal M}_{\bar K^0 K^0}^{(q)}
+ \xi^{\prime}_t{\cal M}_{\bar K^0 K^0}^{(g)},
\eeq
where $\xi_q = V_{qb}V_{qd}^*$, and $\xi^{\prime}_q = V_{qb}V_{qs}^*$ with $q=u,c,t$.

\section{Numerical results}\label{sec:n-d}

In the numerical calculations the following input parameters will be used.
\beq
\Lambda_{\overline{\mathrm{MS}}}^{(5)} &=& 0.225 {\rm GeV},
\quad  f_{B_{s}} = (0.23\pm 0.02) {\rm GeV}, \quad  f_K = 0.16  {\rm GeV},\non
\quad   f_{\pi} &=& 0.13{\rm GeV},\quad
M_{B_{s}} =  5.37 {\rm GeV},\quad   m_K=0.494{\rm GeV},\quad
 m_\eta=547.9{\rm MeV}, \non
 m_\etar&=&0.958{\rm GeV}, \quad m_0^\pi = 1.4 {\rm GeV},\quad   m_0^K = 1.9 {\rm GeV},
\quad \tau_{B_s^0} = 1.470 \ \ {\rm ps},\non
m_b&=&4.8 {\rm GeV}, \quad
M_W = 80.42 {\rm GeV}.
\label{eq:para}
\eeq

For the CKM matrix elements, we also take the same values
as being used in Ref.~\cite{ali07},   and neglect the small errors on
$V_{ud}, V_{us}$, $V_{ts}$ and $V_{tb}$
\beq
|V_{ud}|&=& 0.974, \quad |V_{us}|=0.226,
\quad |V_{ub}|=\left ( 3.68^{+0.11}_{-0.08}\right)\times 10^{-3},\non
\quad |V_{td}|&=&\left ( 8.20^{+0.59}_{-0.27}\right)\times 10^{-3},\quad
|V_{ts}|= 40.96\times 10^{-3}, \quad |V_{tb}|= 1.0,\non
\alpha&=&(99^{+4}_{-9.4})^\circ ,\quad
\gamma=(59.0^{+9.7}_{-3.7})^\circ, \quad
\arg\left [-V_{ts}V_{tb}^* \right] =1^\circ.
\label{eq:angles}
\eeq

\subsection{Branching Ratios}

For the considered $\bar{B}_s^0 \to PP$ decays, the decay amplitude for a given decay mode
with $b \to d$ transition can be generally written as
\beq
{\cal A}(\bar{B}_s^0 \to f) &=& V_{ub}V_{ud}^* T -V_{tb}V_{td}^*  P
= V_{ub}V_{ud}^*T  \left [ 1 + z e^{ i ( -\alpha + \delta ) } \right],
\label{eq:ma}
\eeq
where $\alpha $ is the weak phase (one of the three CKM angles), $\delta=\arg[P/T]$ is the
relative strong phase between the ``T"  and ``P" part, and the parameter  ``z"  is defined as
\beq
 z=\left|\frac{V_{tb} V_{td}^*}{ V_{ub}V_{ud} ^*} \right|
 \left|\frac{P}{T}\right|.
\label{eq:zz}
\eeq
The ratio $z$ and the strong phase $\delta$ can be calculated in the
pQCD approach. The CP-averaged branching ratio, consequently, can be
defined as
\beq
{\rm Br}(B_s^0\to f) = \frac{G_F^2 \tau_{B_s^0}}{32\pi m_B} \;
 \frac{|{\cal A}(\bar{B}_s^0\to f)|^2
+|{\cal A}(B_s^0\to \bar{f})|^2}{2}
 \label{eq:br0}
\eeq
where $\tau_{B_s^0}$ is the lifetime of the $B_s$ meson.

For the case of $b \to s$ transition, we have similarly
\beq
{\cal A}(\bar{B}_s^0 \to f) &=&
V_{ub}V_{us}^* T^\prime  \left [ 1 + z^\prime e^{ i ( \gamma + \delta ) } \right],
\label{eq:ma2}
\eeq
where $\gamma$ is also one of the three CKM angles, $\delta=\arg[P^\prime/T^\prime]$
is the relative strong phase, while the parameter ``$z^\prime$" is defined as
with
\beq
z^\prime=\left|\frac{V_{tb} V_{ts}^*}{ V_{ub}V_{us} ^*} \right|
 \left|\frac{P^\prime}{T^\prime}\right|
\label{eq:zz2}
\eeq

\begin{table}
\caption{ Branching ratios $(\times10^{-6})$ of $B_s \to PP$ decays in the pQCD approach.
The label $LO$ means the leading order pQCD
predictions, while $+VC,+QL,+MP,$ as well as $ NLO$ means that the vertex corrections,
 the quark loops, the magnetic penguins, and all the above $NLO$ corrections are added
 to the $LO$ results, respectively.
 The errors in the table are defined in the context. For comparison, we also cite
 the leading-order pQCD predictions as given in Ref.~\cite{ali07}, the QCDF results in
 Ref.~\cite{npb675}, and currently available data \cite{hfag,npb170,pdg2008}. }
 \label{tab:br}
{\small
\begin{tabular*}{16cm}{@{\extracolsep{\fill}}l|l|ccccc||cc|l} \hline\hline
 Mode & {\rm Class} & {\rm LO} &{\rm + VC}& {\rm + QL}&{\rm + MP} & {\rm NLO}
 & {\rm pQCD}\cite{ali07} & {\rm QCDF}\cite{npb675}& {\rm Data}
 \\ \hline
 $\bar B_s^0\to K^0\eta   $  &$C      $&$0.10  $&$0.08$&$0.12$&$0.10$&$0.19^{+0.03+0.02+0.02}_{-0.06-0.00-0.04}$&$0.11^{+0.08}_{-0.11}$&$0.34^{+0.72}_{-0.33} $&$$\\
 $\bar B_s^0\to K^0\etar  $  &$C      $&$0.70  $&$1.09$&$1.13$&$1.29$&$1.87^{+0.34+0.27+0.13}_{-0.51-0.11-0.21}$&$0.72^{+0.36}_{-0.24}$&$2.0^{+2.2}_{-1.3}  $&$$\\
 $\bar B_s^0\to \pi^0\eta $  &$P_{EW} $&$0.04  $&$0.03$&$-   $&$-   $&$0.03^{+0.01+0.00+0.00}_{-0.00-0.00-0.01}$&$0.05^{+0.02}_{-0.02}$&$0.08^{+0.04}_{-0.03}$&$$\\
 $\bar B_s^0\to \pi^0\etar$  &$P_{EW} $&$0.09  $&$0.08$&$-   $&$-   $&$0.08^{+0.03+0.00+0.01}_{-0.02-0.00-0.01}$&$0.11^{+0.05}_{-0.03}$&$0.11^{+0.05}_{-0.05} $&$$\\
 $\bar B_s^0\to \eta\eta  $  &$P      $&$7.4   $&$7.8 $&$9.1 $&$12.0$&$10.0^{+3.3+0.0+0.6}_{-2.5-0.03-0.8}$&$8.0^{+5.4}_{-3.1} $&$15.6^{+17.0}_{-9.2} $&$$\\
 $\bar B_s^0\to \eta\etar $  &$P      $&$21.6  $&$31.7$&$28.6$&$37.4$&$34.9^{+11.3+0.03+2.8}_{-8.6-0.1-4.1}$&$21.0^{+11.7}_{-7.2}$&$54.0^{+52.8}_{-29.1} $&$$\\
 $\bar B_s^0\to \eta'\etar$  &$P      $&$14.9  $&$30.3$&$18.9$&$27.1$&$25.2^{+8.1+0.1+1.9}_{-6.1-0.0-2.3}$&$14.0^{+7.0}_{-4.1}$&$41.7^{+47.5}_{-24.9} $&$$\\
 $\bar B_s^0\to K^+\pi^-  $  &$T      $&$7.0   $&$6.3 $&$6.1 $&$6.0 $&$6.3^{+2.5+0.5+0.4}_{-1.8-0.5-0.3}  $&$7.6^{+3.3}_{-2.5} $&$10.2^{+6.0}_{-5.2}$&$5.0\pm 1.3   $\\
 $\bar B_s^0\to K^0\pi^0  $  &$C      $&$0.16  $&$0.30$&$0.18$&$0.18$&$0.25^{+0.09+0.03+0.02}_{-0.06-0.01-0.03} $&$0.16^{+0.12}_{-0.07}              $&$0.49^{+0.62}_{-0.35}$&        \\
 $\bar B_s^0\to \pi^+\pi^-$  &$ann    $&$0.70  $&$-   $&$-   $&$-   $&$0.57^{+0.14+0.01+0.20}_{-0.11-0.00-0.19} $&$0.57^{+0.18}_{-0.16}              $&$0.02^{+0.17}_{-0.02}$&$0.53\pm 0.51$\\
 $\bar B_s^0\to \pi^0\pi^0$  &$ann    $&$0.35  $&$-   $&$-   $&$-   $&$0.29^{+0.07+0.01+0.10}_{-0.06-0.00-0.10} $&$0.28^{+0.09}_{-0.08}              $&$0.01^{+0.08}_{-0.01}$&$$\\
 $\bar B_s^0\to K^+K^-    $  &$P      $&$11.8  $&$15.5$&$16.0$&$15.4$&$15.6^{+5.0+0.7+0.8}_{-3.8-0.3-0.7} $&$13.6^{+8.6}_{-5.2}              $&$22.7^{+27.8}_{-13.0}$&$24.4\pm 4.8  $\\
 $\bar B_s^0\to\bar K^0K^0$  &$P      $&$14.3  $&$17.2$&$18.0$&$17.5$&$18.0^{+4.6+0.0+0.7}_{-5.9-0.0-0.6} $&$15.6^{+9.7}_{-6.0}              $&$24.7^{+29.4}_{-14.0}$&        \\
\hline\hline
 \end{tabular*} }
\end{table}

In Table \ref{tab:br} we show the pQCD predictions for the
CP-averaged branching ratios of the thirteen $B_s \to PP$ decays.
The label LO means the leading order pQCD predictions. The label $+$VC, $+$QL, $+$MP,
and NLO means that the vertex corrections,
the quark loops, the magnetic penguins, and all the considered NLO corrections
are included, respectively.
The errors as shown for the NLO pQCD predictions correspond to the uncertainties of the
various input parameters.
The first error comes from $\omega_b=0.50 \pm 0.05$ and $f_{B_s}=0.23\pm 0.02$ GeV.
The second error arises from the uncertainties of  the CKM matrix elements
$|V_{ub}|$ and $|V_{cb}|$, as well as
the CKM angles $\alpha$ and $\gamma$ as given in Eq.~(\ref{eq:angles}).
The first two errors are defined in a similar way as that in Ref.~\cite{ali07}.
The third error comes from the uncertainties of relevant Gegenbauer moments:
$a_1^K=0.17\pm 0.05$, $a_2^K=0.20\pm 0.06$ and $a_2^\pi=0.44^{+0.10}_{-0.20}$.
We here assign roughly a $30\%$ uncertainty for Gegenbauer moments to
estimate the resultant effects on the theoretical predictions of the branching ratios.

For the sake of comparison, we also list the leading order pQCD predictions as given in
Ref.~\cite{ali07} and the theoretical predictions based on the QCD factorization
approach \cite{npb675} in Table \ref{tab:br}.
The corresponding errors of the previous LO pQCD and QCDF predictions denote the
combined error: the individual errors as given in Refs.~\cite{ali07,npb675} are added in
quadrature. The currently available experimental measurements \cite{hfag,npb170,pdg2008}
are also shown in the last column of Table \ref{tab:br}.

From the numerical results about the branching ratios, one can see that
\begin{itemize}

\item
The LO pQCD predictions for branching ratios of $\bar{B}_s \to PP$ decays as given in
Ref.~\cite{ali07} are confirmed by our independent calculation.
The very small differences are induced by the different choices of the scales
$\Lambda_{QCD}^{(4)}$ and $\Lambda_{QCD}^{(5)}$:
we take $\Lambda_{QCD}^{(5)}=0.225$ GeV and the corresponding
$\Lambda_{QCD}^{(4)}=0.287$ GeV, instead of the values of $\Lambda_{QCD}^{(5)}=0.193$ GeV
and $\Lambda_{QCD}^{(4)}=0.25$ GeV as being used in Ref.~\cite{ali07}.

\item
In this paper, the NLO contributions are taken into account partially.
The considered NLO contributions can interfere with the LO part constructively or
destructively for different decay modes. For most decays the changes of the LO results
are moderate and reasonable. The theoretical uncertainty from $\omega_b=0.50\pm 0.05$ is dominant,
while the error from the uncertainty of CKM elements is small. And the total theoretical
error is in general around $30\%$ to $50\%$ in size.

\item
For the ``tree" dominated decay $\bar{B}_s \to K^+ \pi^-$, the NLO
pQCD prediction agrees with the data within one standard deviation.
The agreement between the pQCD prediction and the measured value is improved
due to the inclusion of the considered NLO contribution.

\item
For the three ``Color-suppressed" decays, the NLO enhancement can be significant,
from $\sim 50\%$ for the $\bar{B}_s \to K^0\eta$ and $K^0\pi^0$ decays
to $\sim 170\%$ for the $\bar{B}_s \to K^0\etar$ decay.
The differences between the LO pQCD predictions and the QCDF predictions become
narrow obviously because of the inclusion of the NLO contributions.

\item
For the five ``QCD-Penguin" decays $\bar{B}_s\to \etap \etap$ and $KK$ decays,
the enhancements due to the considered NLO contributions can be as large as
$(30-70)\%$, which are helpful to pin down the gap between the pQCD and the QCDF
predictions.

\item
For the two ``Electroweak-Penguin" decays $\bar{B}_s \to \pi^0 \etap$, the NLO
contributions are small. The pQCD predictions agree well with the QCDF predictions.

\item
For the ``annihilation" decays $\bar{B}_s \to \pi^+\pi^-$ and $\pi^0 \pi^0$ decays,
the NLO contributions are around $10\%$ only. The pQCD predictions agree well with
the measured value.

\item
For the considered thirteen $B_s \to PP$ decays, only
three of them, $B_s \to K^+\pi^-, K^+ K^-$ and $\pi^+ \pi^-$, have been measured experimentally
with good precision.
It is easy to see that the consistency between the pQCD predictions for their branching ratios
and the measured values will be improved effectively when the NLO contributions are included.

\end{itemize}

\subsection{CP-violating asymmetries}

\begin{table}
\caption{Direct CP asymmetries(in $\%$) of $B_s \to PP$ decays in the pQCD approach.
The label $LO$ means the leading order pQCD
predictions, while $+VC,+QL,+MP,$ as well as $ NLO$ means that the vertex corrections,
 the quark loops, the magnetic penguins, and all the above $NLO$ corrections are added
 to the $LO$ results, respectively.
 The errors in the table are defined in the context. For comparison, we also cite
 the leading-order pQCD predictions as given in Ref.~\cite{ali07}, the QCDF results in
 Ref.~\cite{npb675}.}
\label{tab:acp1}
\begin{tabular*}{16cm}{@{\extracolsep{\fill}}l|l|ccccc|cc} \hline\hline
 Mode & {\rm Class} & {\rm LO} &{\rm + VC}& {\rm + QL}&{\rm + MP} & {\rm NLO}
 & {\rm pQCD}\cite{ali07} & {\rm QCDF}\cite{npb675}  \\ \hline
 $\bar B_s^0\to K^0\eta   $  &$C      $&$65.0  $&$45.4 $&$38.6 $&$30.7$&$96.7^{+0.0+1.1+1.2}_{-0.1-2.0-1.5}$&$56.4^{+8.0}_{-9.3}$&$46.8^{+48.8}_{-58.8} $\\
 $\bar B_s^0\to K^0\etar  $  &$C      $&$-22.3 $&$-5.7 $&$-18.1$&$-0.8$&$-35.4^{+2.0+2.4+0.5}_{-0.0-2.5-0.3}  $&$-19.9^{+5.5}_{-5.3}$&$-36.6^{+22.3}_{-20.7}  $\\
 $\bar B_s^0\to \pi^0\eta $  &$P_{EW} $&$0.3   $&$40.4 $&$-    $&$-   $&$40.4^{+0.3+1.6+3.6}_{-0.8-1.3-7.2}   $&$-0.4^{+0.3}_{-0.3} $&$-$\\
 $\bar B_s^0\to \pi^0\etar$  &$P_{EW} $&$23.8  $&$52.5 $&$-    $&$-   $&$52.5^{+2.0+2.4+0.5}_{-0.0-2.5-0.3}   $&$20.6^{+3.4}_{-2.9} $&$27.8^{+27.2}_{-28.8} $\\
 $\bar B_s^0\to \eta\eta  $  &$P      $&$-0.7  $&$-1.53$&$1.2  $&$1.1 $&$0.6^{+0.1+0.1+0.2}_{-0.0-0.0-0.0}    $&$-0.6^{+0.6}_{-0.5} $&$-1.6^{+2.4}_{-2.4} $\\
 $\bar B_s^0\to \eta\etar $  &$P      $&$-1.3  $&$-1.1 $&$-0.2 $&$-0.6$&$-0.2^{+0.1+0.0+0.1}_{-0.1-0.0-0.1}    $&$-1.3^{+0.1}_{-0.2} $&$0.4^{+0.5}_{-0.4} $\\
 $\bar B_s^0\to \eta'\etar$  &$P      $&$1.9   $&$1.3  $&$1.0  $&$1.3 $&$1.4^{+0.1+0.1+0.1}_{-0.1-0.1-0.2}    $&$1.9^{+0.4}_{-0.5}  $&$2.1^{+1.3}_{-1.4} $\\
 $\bar B_s^0\to K^+\pi^-  $  &$T      $&$25.7  $&$28.6 $&$25.7 $&$28.5 $&$25.8^{+4.1+1.5+2.7}_{-3.8-0.7-5.0}  $&$24.1^{+5.6}_{-4.8}$&$-6.7^{+15.6}_{-15.3}$\\
 $\bar B_s^0\to K^0\pi^0  $  &$C      $&$66.9  $&$86.4 $&$-18.7$&$-10.8$&$88.0^{+3.7+2.6+1.6}_{-4.5-6.7-1.6}  $&$59.4^{+7.9}_{-12.5}              $&$42^{+47}_{-56}$        \\
 $\bar B_s^0\to \pi^+\pi^-$  &$ann    $&$-1.1  $&$-    $&$-    $&$-    $&$0.2^{+0.1+0.0+2.0}_{-0.0-0.0-1.5}   $&$-1.2^{+1.2}_{-1.3}              $&$-$\\
 $\bar B_s^0\to \pi^0\pi^0$  &$ann    $&$-1.1  $&$-    $&$-    $&$-    $&$0.2^{+0.1}_{-1.5}   $&$-1.2^{+1.2}_{-1.2}              $&$-$\\
 $\bar B_s^0\to K^+K^-    $  &$P      $&$-22.1 $&$-17.9$&$-14.1$&$-17.1$&$-15.6^{+1.2+0.7+1.3}_{-0.8-0.9-1.1} $&$-23.3^{+5.0}_{-4.6}             $&$4.0^{+10.6}_{-11.6}$\\
 $\bar B_s^0\to\bar K^0K^0$  &$P      $&$0     $&$0    $&$0.3  $&$0    $&$0.4\pm 0.1   $&$0                 $&$0.9\pm0.4$        \\
\hline\hline
\end{tabular*}
\end{table}

Now we turn to the evaluations of the CP-violating asymmetries of
$B_{s} \to PP$ decays in pQCD approach.
Restricting the final state f to have definite CP, the
time-dependent decay width for the $B_s \to f$ decay can be written as
~\cite{prd52}
\beq
\Gamma(\bar{B}_s^0(t)\to f)&=& e^{-\Gamma t}\; \overline{\Gamma}\left (\bar{B}^0_s \to
f\right)\cdot \left [ \cosh\left (\frac{\Delta \Gamma t}{2}\right )
\right.\non
&&\left.
+H_f\sinh\left (\frac{\Delta \Gamma t}{2}\right )
+{\cal A}_{CP}^{dir}\cos(\Delta m\, t) + S_f\sin(\Delta m\, t)\right ]
\eeq
where $\Delta m=m_H-m_L>0$, $\overline{\Gamma}=(\Gamma_H+\Gamma_L)/2)$
is the average decay
widths, while $\Delta \Gamma =\Gamma_H-\Gamma_L $ is the difference of
decay widths for the heavier and lighter $B_s^0$ mass eigenstates.
In the $B_s$ system, we expect a much larger decay width difference:
$(\Delta\Gamma/\Gamma )_{B_s}\sim -20\% $ \cite{hfag}.
Besides $\cala^{dir}$, the CP-violating asymmetry  $S_f$ and $H_f$ can be
defined as
\beq
\acp^{dir}=\frac{|\lambda|^2-1 }{1+|\lambda|^2},\quad
S_f=\frac{2 {\rm Im}[\lambda]}{1+|\lambda|^2},\quad
H_f=\frac{2 {\rm Re}[\lambda]}{1+|\lambda|^2},
\eeq
with the parameter $\lambda$
\beq
\lambda=\eta_f e^{2i\epsilon}\frac{A(\bar B_s \to f)}{A(B_s \to \bar f)},
\eeq
where $\eta_f$ is $+1(-1)$ for a CP-even(CP-odd) final state
f and $\epsilon =\arg[-V_{ts}V_{tb}^*]$ is very small in size.

If we neglect the very small parameter $\epsilon$, the CP-violating asymmetries can be written
explicitly as
\beq
{\cal A}_{CP}^{dir} &=&
\frac{2 z \sin \alpha \sin\delta}{1+2 z\cos \alpha \cos \delta +z^2},
\label{eq:acp011}, \non
S_f&=&-\frac{\sin(2\gamma)+z^2\sin(2\gamma+2\alpha)+2z\cos\delta
\sin(\alpha+2\gamma)}{1+z^2+2z\cos\delta \cos\alpha},
\label{eq:acp012}\non
H_f&=&\frac{2z\cos(\delta)\cos(\alpha+2\gamma)+\cos(2\gamma)+
z^2\cos(2\alpha+2\gamma)}{1+z^2+2z\cos\delta \cos\alpha},
\label{eq:acp013}
\eeq
for the decays relevant to the $b \to d $ transition, and
\beq
{\cal A}_{CP}^{dir} &=& -\frac{2 z^\prime \sin\gamma \sin\delta}{1
+2 z^\prime \cos\gamma \cos\delta +z^{\prime 2}},\label{eq:acp021}, \non
S_f&=& -\frac{\sin(2\gamma)+2 z^\prime \cos\delta
\sin\gamma}{1+z^{\prime 2}+2z^\prime \cos\delta\cos\gamma},
\label{eq:acp022}\non
H_f&=& \frac{z^{\prime 2}+2z^\prime \cos\delta \cos\gamma
+\cos(2\gamma)}{1+z^{\prime 2}+2z^\prime \cos\delta \cos\gamma},
\label{eq:acp023}
\eeq
for the case of $b \to s$ transition.

The pQCD predictions for the direct CP asymmetries $\cala^{dir}$, the
mixing-induced CP asymmetries $S_f$ and $H_f$ of the considered $B_s^0\to PP$ decays
are listed in Table \ref{tab:acp1} and Table \ref{tab:acp2}.
In both tables, the label LO means the leading order pQCD predictions, and the labels
$+$VC, $+$QL, $+$MP, as well as NLO mean that the vertex corrections,
the quark loops, the magnetic penguins, and all the above NLO
corrections are added to LO results, respectively.
As a comparison, the LO pQCD predictions as given in Ref.~\cite{ali07} are
listed in Table \ref{tab:acp1} and Table \ref{tab:acp2}.
In Table \ref{tab:acp1}, the QCDF predictions for direct CP-violating asymmetries as
given in Ref.~\cite{npb675} are also shown.
The corresponding errors of the previous LO pQCD predictions and QCDF predictions
are the combined errors: the individual errors as given in Refs.~\cite{ali07,npb675}
are added in quadrature.
The errors of our NLO pQCD predictions for CP-violating asymmetries
are defined in the same way as those for the branching ratios.

For the experimental measurements, there is only one measured CP asymmetry as reported by CDF
Collaboration \cite{npb170}:
\beq
\acp^{dir}(\bar{B}_s\to K^+ \pi^-)=0.39\pm 0.17.
\label{eq:acp-d1}
\eeq
But more data will become available soon when the LHC starts its physics running.

From the pQCD predictions and currently available experimental measurements for the CP violating
asymmetries of the thirteen $B \to PP$ decays, one can see the following points:
\begin{itemize}
\item
The LO pQCD predictions obtained in this paper agree very well with those as given in
Ref.~\cite{ali07}. For $\bar{B}_s \to K^0 \eta$ and $\pi^0 \eta$ decays, the LO pQCD
predictions can be changed significantly by the inclusion of the NLO contributions.
For other decays, the NLO contributions are small or moderate in size. The pQCD predictions
are in general consistent with those in QCDF approach, but much larger than the later one for
$\bar{B}_s \to K^\eta, \pi^0\etar$ and $K^0\pi^0$ decays.

\item
For the ``Tree" dominated decay $\bar{B}_s \to K^+\pi^-$, the pQCD prediction for the
direct CP asymmetry is $\acp^{dir}(\bar{B}_s \to K^+\pi^-)=0.30\pm 0.06$, which
agrees very well with the experimental measurement as given in Eq.~(\ref{eq:acp-d1}).
The QCDF prediction, however, is about $-0.07\pm 0.16$ and much different
from the measured value.

\item
For the four ``QCD-penguin" decays $\bar{B}_s \to \etap \etap$ and $\bar{K}^0 K^0$ decays,
analogous to the QCDF predictions, the LO and NLO pQCD predictions for both $\acp^{dir}$
and $S_f$ are all very small in size.

\item
For the ``annihilation" decays $\bar{B}_s \to \pi^+\pi^-$ and $\pi^0\pi^0$, the pQCD predictions
for the direct CP-violating asymmetries are very small in size, while $S_f$ is around $10\%$ and
$H_f\sim 1$.

\end{itemize}

\begin{table}
\caption{The mixing-induced CP asymmetries (in $\%$) $S_f$ and $H_f$ (the second row). The
label LO means the leading order pQCD predictions, and the lables
$+$VC, $+$QL,$+$MP, as well as NLO mean that the vertex corrections,
 the quark loops, the magnetic penguins, and all the above NLO
 corrections are added to LO results, respectively.
 The errors  of the entries are defined in the context. As a comparison, the LO pQCD predictions
 as given in Ref.~\cite{ali07} are also listed.}
\label{tab:acp2}
\begin{tabular}{l |l|c c c c c|c } \hline \hline
 Mode &Class & LO & +VC & +QL &+MP &NLO & pQCD\cite{ali07} \\ \hline
$\bar B_s^0\to K_s^0\eta   $&$C$&$-37  $&$-89 $&$-92 $&$-90$&$-18^{+0+7+3}_{-2-11-5}$&$-43^{+23}_{-23}$\\
                            &$$&$-67  $&$-57 $&$6   $&$32 $&$-18^{+2+18+15}_{-0-8-9}$&$-70^{+14}_{-22}$\\
$\bar B_s^0\to K_s^0\etar  $&$C$&$-67  $&$-59 $&$-44 $&$-53$&$-46^{+1+12+1}_{-0-23-0}$&$-68^{+6}_{-5}$\\
                            &$$&$-70  $&$-80 $&$-88 $&$-85$&$-82^{+0+20+1}_{-1-6-0}$&$-70^{+6}_{-7}$\\ \hline
$\bar B_s^0\to\pi^0\eta    $&$P_{EW}$&$ 18  $&$28  $&$-   $&$-  $&$28^{+2+3+4}_{-3-1-4} $&$17^{+18}_{-13}$\\
                            &$$&$ 98  $&$87  $&$-   $&$-  $&$87^{+1+1+4}_{-1-1-2} $&$99^{+1}_{-2} $\\
$\bar B_s^0\to\pi^0\etar   $&$P_{EW}$&$-25  $&$-18 $&$-   $&$-  $&$-18^{+1+12+1}_{-0-23-0}$&$-17^{+8}_{-9}$\\
                            &$$&$ 94  $&$83  $&$-   $&$-  $&$83^{+3+17+1}_{-1-1-0} $&$96^{+2}_{-2} $\\ \hline
$\bar B_s^0\to\eta\eta     $&$P$&$ 3   $&$0   $&$1   $&$1  $&$2^{+0+0+0}_{-0-0-0}  $&$3^{+1}_{-1}  $\\
                            &$$&$ 100 $&$100 $&$100 $&$100$&$100^{+0+0+0}_{-0-0-0}$&$100^{+0}_{-0}$\\
$\bar B_s^0\to\eta\etar    $&$P$&$ 4   $&$3   $&$4   $&$3  $&$4^{+0+0+0}_{-0-0-0}  $&$4^{+0}_{-0}  $\\
                            &$$&$ 100 $&$100 $&$100 $&$100$&$100^{+0+0+0}_{-0-0-0}$&$100^{+0}_{-0}$\\
$\bar B_s^0\to\etar\etar   $&$P$&$ 4   $&$6   $&$6   $&$6  $&$5^{+0+0+0}_{-1-1-0}  $&$4^{+1}_{-1}  $\\
                            &$$&$ 100 $&$100 $&$100 $&$100$&$100^{+0+0+0}_{-0-0-0}$&$100^{+0}_{-0}$\\ \hline
$\bar B_s^0\to K_s^0\pi^0$  &$C$&$-55 $&$-25$&$-98 $&$-97$ &$-41^{+8+4+3}_{-9-8-5}             $&$-61^{+24}_{-20} $  \\
                            &$ $&$-50 $&$-44$&$-8  $&$-20$ &$-23^{+0+19+3}_{-1-18-7}            $&$-52^{+23}_{-17} $  \\
$\bar B_s^0\to K^+K^-$      &$P$&$24  $&$20 $&$22  $&$20 $ &$22^{+2+2+2}_{-2-1-2}               $&$28 ^{+5}_{-5} $  \\
                            &$$&$95  $&$96 $&$96  $&$96 $ &$96^{+4+0+0}_{-3-1-0}               $&$93 ^{+3}_{-3} $  \\
$\bar B_s^0\to \bar{K}^0K^0$&$P$&$ -  $&$-  $&$0.4 $&$-$   &$0.4^{+0+0+0}_{-0-0-0}              $&$4             $  \\
                            &$$&$ -  $&$-  $&$100 $&$-$   &$100^{+0+0+0}_{-0-0-0}              $&$100            $ \\ \hline
$\bar B_s^0\to \pi^+\pi^-$  &$ann$&$9.5 $&$-  $&$-   $&$-$   &$9^{+1+1+1}_{-0-0-0}  $&$14 ^{+12}_{-6}$  \\
                            &$$&$99.5$&$-  $&$-   $&$-$   &$100^{+0+0+0}_{-0-0-0} $&$99 ^{+0}_{-1}$   \\
$\bar B_s^0\to \pi^0\pi^0$  &$ann$&$9.5 $&$-  $&$-   $&$-$   &$8.1^{+0.1+0.5+0.3}_{-0.3-0.7-0.0}   $    &$14 ^{+12}_{-6}$  \\
                            &$$&$99.5$&$-  $&$-   $&$-$   &$100^{+0+0+0}_{-0-0-0}  $    &$99 ^{+0}_{-1}$   \\
\hline\hline
\end{tabular}
\label{tab:diracp}
\end{table}

\section{SUMMARY}

In this paper, we calculated the partial NLO contributions to the
branching ratios and CP-violating asymmetries of $\bar B_s^0\to PP$ decays.
Here the NLO contributions from the QCD vertex corrections,
the quark-loops and the chromo-magnetic penguins are included.

From our calculations and phenomenological analysis, we found the following results:
\begin{itemize}
\item
The LO pQCD predictions for the branching ratios and CP-violating asymmetries of
$B_s \to PP$ decays as presented in Ref.~\cite{ali07} are confirmed
by our independent calculation.

\item
For branching ratios, the effects of the considered NLO contributions
are varying from small to significant for different decay mode.
For the three measured decays $\bar{B}_s \to K^+ \pi^-, K^+K^-$ and $\pi^+ \pi^-$,
for example, the consistency between the pQCD predictions and the measured values are
improved effectively due to the inclusion of the considered NLO contributions.
For the three ``Color-suppressed" decays, for instance, the NLO enhancement can be significant,
from $\sim 50\%$ for the $\bar{B}_s \to K^0\eta$ and $K^0\pi^0$ decays
to $\sim 170\%$ for the $\bar{B}_s \to K^0\etar$ decay, to be tested by forthcoming
LHC experiments.

\item
As for the CP-violating asymmetries, the LO pQCD predictions for
$\bar{B}_s \to K^0 \eta$ and $\pi^0 \eta$ decays could be changed significantly by
the inclusion of the NLO contributions.
For other decays, the NLO contributions are small or moderate in size.
For $\bar{B}_s \to K^+\pi^-$ decay, the pQCD prediction for the
direct CP asymmetry is $\acp^{dir}(\bar{B}_s \to K^+\pi^-)=0.26\pm 0.06$, which
agrees very well with the measured value $\acp^{dir}(\bar{B}_s \to K^+\pi^-)=0.39\pm 0.17$.

\item
In this paper, only the partial NLO contributions in the pQCD approach
have been taken into account. The still missing pieces
relevant with the emission diagrams, hard-spectator and annihilation diagrams
should be evaluated as soon as possible.

\end{itemize}

\begin{acknowledgments}
The authors would like to thank Cai-Dian L\"u, Li-bo Guo, Zhi-qing Zhang, Xin Liu
and Po Li for helpful discussions.
This work is partly supported by the National Natural Science
Foundation of China under Grant No.10575052, 10605012 and 10735080.

\end{acknowledgments}


\begin{appendix}

\section{Related Functions }\label{sec:aa}

We show here the hard function $h_i$ and the Sudakov factors $S_{ab,cd,ef,gh}(t)$
appeared in the expressions of the decay amplitudes in Sec.~\ref{sec:lo} and \ref{sec:nlo}.
The hard functions $h_i(x_j,b_j)$ are obtained by making the Fourier
transformations  of the hard kernel $H^{(0)}$.
\beq
h_e(x_1,x_3,b_1,b_3)&=& \left [\theta(b_1-b_3)I_0\left(\sqrt{x_3}
M_{B_s}b_3 \right )K_0\left (\sqrt{x_3} M_{B_s}b_1\right )
 +\theta(b_3-b_1)I_0\left (\sqrt{x_3}  M_{B_{s}} b_1\right )
 \right.\non
 && \left.
\times K_0\left (\sqrt{x_3}M_{B_{s}} b_3 \right )\right ]
K_0\left (\sqrt{x_1 x_3} M_{B_{s}} b_1\right ) S_t(x_3),
\label{he1}
\eeq
\beq
 h_n(x_i,b_1,b_2) &=&
 \left [ \theta(b_2-b_1) \mathrm{K}_0(M_{B_s}\sqrt{x_1 x_3} b_2)
 \mathrm{I}_0(M_{B_s}\sqrt{x_1 x_3} b_1)
\right. \non
&& \left.
+ \theta(b_1-b_2) \mathrm{K}_0(M_{B_s}\sqrt{x_1 x_3} b_1)
 \mathrm{I}_0(M_{B_s}\sqrt{x_1 x_3} b_2)\right ]
 \non && \times
\left\{\begin{array}{ll}
 \frac{i\pi}{2} \mathrm{H}_0^{(1)}\left (M_{B_s}\sqrt{(x_2-x_1)x_3}b_2\right ),&
 \textrm{for}\quad x_1-x_2<0,\\
 \mathrm{K}_0\left (M_{B_s}\sqrt{(x_2-x_1)x_3}b_2 \right ), & \textrm{for}\quad
 x_2-x_1>0, \end{array}
\right. \label{eq:pp1}
\eeq

\beq
 h_a(x_2,x_3,b_2,b_3) &=&
 \left [\theta(b_2-b_3)\mathrm{K}_0\left (i \sqrt{x_3} M_{B_s} b_2\right )
 \mathrm{I}_0\left (i \sqrt{x_3} M_{B_{s}} b_3\right )
 +\theta(b_3-b_2)
 \right.\non
&&\left.
\times\mathrm{K}_0\left (i \sqrt{x_3} M_{B_s} b_3\right )
\mathrm{I}_0\left (i\sqrt{x_3} M_{B_{s}} b_2\right ) \right ]
\mathrm{K}_0\left (i \sqrt{x_2 x_3} M_{B_s} b_2\right ) S_t(x_3), \label{he3}
\eeq

\beq
 h_{na}(x_i,b_1,b_2) &=&
 \left [\theta(b_1-b_2) \mathrm{K}_0\left (i \sqrt{x_2(1- x_3)}M_{B_s} b_1\right )
 \mathrm{I}_0\left (i \sqrt{x_2(1- x_3)} M_{B_s} b_2\right )
 \right.\non
 &&\left.
+ \theta(b_2-b_1) \mathrm{K}_0\left (i \sqrt{x_2(1- x_3)}M_{B_s} b_2\right )
 \mathrm{I}_0\left (i \sqrt{x_2(1- x_3)} M_{B_s}  b_1\right )\right ]
 \non && \times\mathrm{K}_0\left (\sqrt{1-(1-x_1-x_2)x_3}M_{B_s}b_2\right ),
\label{eq:pp3}
\eeq

\beq
 h'_{na}(x_i,b_1,b_2) &=&
\left [\theta(b_1-b_2) \mathrm{K}_0\left (i \sqrt{x_2(1- x_3)}M_{B_s} b_1\right )
 \mathrm{I}_0\left (i \sqrt{x_2(1- x_3)} M_{B_s} b_2\right )\right.
 \non
 && \left.
 + \theta(b_2-b_1) \mathrm{K}_0\left (i \sqrt{x_2(1- x_3)}M_{B_s} b_2\right )
 \mathrm{I}_0\left (i \sqrt{x_2(1- x_3)} M_{B_s}
 b_1\right )\right ]
 \non && \times\left\{
\begin{array}{ll}
 \frac{i\pi}{2} \mathrm{H}_0^{(1)}\left (M_{B_s}\sqrt{(x_2-x_1)(1-x_3)}b_1\right ),&
 \textrm{for}\quad x_1-x_2<0,\\
 \mathrm{K}_0\left (M_{B_s}\sqrt{(x_2-x_1)(1-x_3)}b_1 \right ),
 & \textrm{for}\quad  x_2-x_1>0,
\end{array}\right.
 \label{eq:pp4}
\eeq

\beq
h_g(x_i,b_i)&=&-\frac{i\pi}{2}S_t(x_3)
\left [J_0\left (\sqrt{x_2\bar{x}_3}M_{B_s}b_2\right )
+iN_0\left (\sqrt{x_2\bar{x}_3}M_{B_s}b_2\right )\right ]
K_0\left (\sqrt{x_1x_3}M_{B_s}b_1\right )
\non &&
\cdot \int_0^{\pi/2}\!d\theta \tan{\theta}
\cdot J_0\left(\sqrt{x_3}M_{B_s}b_1\tan{\theta}\right )
J_0\left (\sqrt{x_3}M_{B_s}b_2\tan{\theta}\right )
\non &&
\hspace{2cm}\cdot J_0\left (\sqrt{x_3}M_{B_s}b_3\tan{\theta}\right ),
\eeq

\beq
h'_g(x_i,b_i)&=& -S_t(x_1) K_0\left (\sqrt{x_1x_3}M_{B_s}b_3 \right )
\cdot \int_0^{\pi/2}\!d\theta \tan{\theta}\cdot
J_0\left (\sqrt{x_1}M_{B_s}b_1\tan{\theta}\right )
\non &&
\ \ \ \ \cdot J_0\left (\sqrt{x_1}M_{B_s}b_2\tan{\theta} \right )
\; J_0\left (\sqrt{x_1}M_{B_s}b_3\tan{\theta} \right )\non
&& \ \ \ \ \times
\left\{ \begin{array}{ll}\frac{i\pi}{2}
\left [ J_0\left (\sqrt{x_2-x_1}M_{B_s}b_2 \right )
+iN_0\left (\sqrt{x_2-x_1}M_{B_s}b_2\right )\right ],&
x_1<x_2,\\
K_0\left (\sqrt{x_1-x_2}M_{B_s}b_2 \right ),& x_1>x_2,\end{array}\right.
\eeq
where
\beq
K_0 (i x) =\frac{i\pi}{2}\mathrm{H}_0^{(1)}(x)= \frac{i\pi}{2}
\left [ iY_0 (x) + J_0 (x) \right],
\eeq
with $K_0$, $I_0$ and $J_0$ are the Bessel functions~\cite{isg}. And the threshold
resummation form factor $S_t(x_i)$ can be found in Ref.~\cite{tk07}.

The Sudakov factors appeared in Eqs.~(\ref{eq:ab}-\ref{eq:gh}) are
defined as
\beq
S_{ab}(t) &=& s\left ( x_1\frac{m_{B_s}}{\sqrt 2},\,b_1 \right )
+ s\left (x_3\frac{m_{B_s}}{\sqrt{2}},\,b_3 \right )
+ s\left (\bar{x}_3\frac{m_{B_s}}{\sqrt 2},\,b_3 \right )
\non &&+\frac{5}{3}\int_{1/b_1}^t d\mu
\frac{\gamma_q(\alpha_s(\mu))}{\mu}+ 2\int_{1/ b_3}^t d\mu
\frac{\gamma_q(\alpha_s(\mu))}{\mu},
\label{Sa}
\eeq
\beq
S_{cd}(t) &=& s\left ( x_1\frac{m_{B_s}}{\sqrt 2},\,b_1 \right )
+ s\left ( x_2\frac{m_{B_s}}{\sqrt{2}},\,b_2 \right )
+ s\left (\bar{x}_2 \frac{m_{B_s}}{\sqrt2},\,b_2\right )
+ s\left (x_3\frac{m_{B_s}}{\sqrt{2}},\,b_1 \right )
\non &&
+ s\left (\bar{x}_3\frac{m_{B_s}}{\sqrt 2},\,b_1\right )
+\frac{11}{3}\int_{1/b_1}^t d\mu
\frac{\gamma_q(\alpha_s(\mu))}{\mu}+ 2\int_{1/ b_2}^t d\mu
\frac{\gamma_q(\alpha_s(\mu))}{\mu},
\label{Sc}
\eeq

\beq
S_{ef}(t) &=& s\left (x_2\frac{m_{B_s}}{\sqrt 2},\,b_2 \right )
+ s\left (\bar{x}_2\frac{m_{B_s}}{\sqrt 2},\,b_2\right )
+s\left (x_3\frac{m_{B_s}}{\sqrt{2}},\,b_3\right )
\non &&
+ s\left (\bar{x}_3\frac{m_{B_s}}{\sqrt 2},\,b_3 \right )
+ 2\int_{1/b_2}^t
d\mu \frac{\gamma_q(\alpha_s(\mu))}{\mu}+ 2\int_{1/ b_3}^t d\mu
\frac{\gamma_q(\alpha_s(\mu))}{\mu},
\label{Se}
\eeq

\beq
S_{gh}(t) &=& s\left ( x_1\frac{m_{B_s}}{\sqrt 2},\,b_1 \right )
+ s\left (x_2\frac{m_{B_s}}{\sqrt{2}},\,b_2\right )
+s\left (\bar{x}_2\frac{m_{B_s}}{\sqrt2},\,b_2\right )
+s\left (x_3\frac{m_{B_s}}{\sqrt{2}},\,b_2\right )
\non &&
+ s\left (\bar{x}_3\frac{m_{B_s}}{\sqrt 2},\,b_2\right )
+ \frac{5}{3}\int_{1/b_1}^t d\mu
\frac{\gamma_q(\alpha_s(\mu))}{\mu}+ 4\int_{1/ b_2}^t d\mu
\frac{\gamma_q(\alpha_s(\mu))}{\mu} \label{Sg},
\eeq
where the quark
anomalous dimension $\gamma_q=-\alpha_s/ \pi$ and the function
$s(Q,b)$ is given as \cite{plb555,epjc695}:
\beq
s(Q,b)=\int_{1/b}^Q\; \frac{d\mu}{\mu}\Bigl[
\ln\left(\frac{Q}{\mu}\right)A(\alpha(\bar\mu))+B(\alpha_s(\bar\mu))
\Bigr] \label{su1}
\eeq
with
\beq
A&=&\frac{4}{3}\frac{\alpha_s}{\pi}+\left[\frac{67}{9}-\frac{\pi^2}{3}-\frac{10}{27}N_{f}+
\frac{2}{3}\beta_0\ln\left(\frac{e^{\gamma_E}}{2}\right)\right]
 \left(\frac{\alpha_s}{\pi}\right)^2 ,\non
B&=&\frac{2}{3}\frac{\alpha_s}{\pi}\ln\left(\frac{e^{2\gamma_{E}-1}}{2}\right),
\eeq
where $\gamma_E=0.57722\cdots$ is the Euler constant, an $N_{f}$
is the number of active quark flavors.
The hard scales $t_e^i$ appeared in the above equations take the form of
\beq
t_{e}^1 &=& {\rm max}\left \{\sqrt{x_3} M_{B_s},1/b_1,1/b_3\right \},\non
t_{e}^2 &=& {\rm max}\left\{\sqrt{x_1} M_{B_s},1/b_1,1/b_3\right\},\non
t_{e}^3 &=& {\rm max}\left\{\sqrt{x_1x_3}M_{B_s},
         \sqrt{\mid1-x_1-x_2\mid x_3} M_{B_s},1/b_1,1/b_2\right\},\non
t_{e}^4 &=& {\rm max}\left\{\sqrt{x_1x_3} M_{B_s},
         \sqrt{\mid x_1-x_2\mid x_3} M_{B_s},1/b_1,1/b_2\right\},\non
t_{e}^5 &=& {\rm max}\left\{\sqrt{\bar{x}_3}M_{B_s},1/b_2,1/b_3\right\},\non
t_{e}^6 &=& {\rm max}\left\{\sqrt{x_2}M_{B_s},  1/b_2,1/b_3\right\}, \non
t_{e}^7 &=& {\rm max}\left\{\sqrt{x_2\bar{x}_3}M_{B_s},
         \sqrt{1-(1-x_1-x_2)x_3} M_{B_s},1/b_1,1/b_2 \right\},\non
t_{e}^8 &=& {\rm max}\left \{\sqrt{x_2\bar{x}_3} M_{B_s},
         \sqrt{\mid x_1-x_2\mid \bar{x}_3} M_{B_s},1/b_1,1/b_2 \right\}.
\eeq
They are chosen as the maximum energy scale appearing in each diagram
to kill the large logarithmic radiative corrections.

\end{appendix}




\begin{thebibliography}{99}

\bibitem{lhcb1}
G.~Buchalla {\it et al.},  \epjc {\bf 57}, 309(2008) and references therein.

\bibitem{aag}
A.~Ali, G.~Kramer and C.D.~L\"u, \prd {\bf 58}, 094009 (1998);
{\it ibid.} {\bf 59}, 014005 (1998).

\bibitem{yhc}
Y.-H.~Chen, H.Y.~Cheng, B.~Tseng, and  K.C.~Yang, \prd {\bf 60}, 094014 (1999);
H.Y.~Cheng and K.C.~Yang, \prd {\bf 62}, 054029  (2000).

\bibitem{bbns99}
M.~Beneke, G.~Buchalla, M.~Neubert and C.T.~Sachrajda,
\prl {\bf 83}, 1914 (1999); \npb {\bf 591}, 313 (2000).

\bibitem{chenbs99}
Y.H.~Chen, H.Y.~Cheng, B.~Tseng,  \prd {\bf 59}, 074003 (1999).

\bibitem{npb675}
M.~Beneke and M.~Neubert, \npb {\bf 675}, 333 (2003).

\bibitem{xiaobs01}
D.~Zhang, Z.J.~Xiao, and C.S.~Li, \prd {\bf 64}, 014014 (2001).

\bibitem{pipi}
Y.~Li, C.D.~L\"u, Z.J.~Xiao, and X.Q.~Yu,  \prd {\bf 70}, 034009 (2004);
X.Q.~Yu, Y.~Li, and C.D.~L\"u, \prd {\bf 71}, 074026 (2005); \prd {\bf 73}, 017501 (2006);
J.~Zhu, Y.L.~Shen, and C.D.~L\"u, \jpg {\bf 32}, 101  (2006).

\bibitem{pieta}
Z.J. ~Xiao, X.~Liu and H.S.~Wang, \prd {\bf 75}, 034017 (2007).

\bibitem{ali07}
A.~Ali, G.~Kramer, Y.~Li, C.D.~L\"u, Y.L.~Shen, W.~Wang and Y.M.~Wang,
\prd {\bf 76}, 074018 (2007).

\bibitem{nlo05}
H.N.~Li, S.~Mishima, A.I.~Sanda, \prd {\bf 72}, 114005 (2005).

\bibitem{xiao08a}
Z.Q.~Zhang and Z.J.~Xiao, arXiv: 0807.2022 [hep-ph]; arXiv: 0807.2024 [hep-ph];

\bibitem{xiao08b}
Z.J.~Xiao, Z.Q.~Zhang,  X.~Liu, and L.B.~Guo, \prd {\bf 78}, 114001 (2008).

\bibitem{li2003}
H.N.~Li,  Prog. Part. $\&$ Nucl. Phys. {\bf 51}, 85 (2003) and references therein.

\bibitem{buras96}
G.~Buchalla, A.J.~Buras, M.E.~Lautenbacher, \rmp {\bf 68}, 1125 (1996).

\bibitem{scet01}
C.W.~Bauer, D.~Pirjol, I.Z.~Rothstein and I.W.~ Stewart, \prd {\bf 70}, 054015 (2004);
M.~Beneke and D.S.~Yang, \npb {\bf 736}, 34 (2006).

\bibitem{fks98}
T.~Feldmann, P.~Kroll, and B.~Stech, \prd {\bf 58}, 114006 (1998);
T.~Feldmann, Int. J. Mod. Phys. A{\bf 15}, 159 (2000).

\bibitem{pball98}
V.M.~Braun and I.E.~Filyanov , \zpc {\bf 48}, 239 (1990);
P.~Ball, V.M.~Braun, Y.~Koike, and K.~Tanaka, \npb {\bf 529}, 323 (1998);
P.~Ball, \jhep {\bf 01}, 010 (1999).

\bibitem{pball06}
V.M.~Braun and A.~Lenz, \prd {\bf 70}, 074020 (2004);
P.~Ball and A.~talbot, \jhep {\bf 06}, 063 (2005);
P.~Ball and R.~Zwicky, \plb {\bf 633}, 289 (2006);
A.~Khodjamirian, Th.~Mannel, and M.~Melcher, \prd {\bf 70}, 094002 (2004).

\bibitem{o8g2003}
S.~Mishima and A.I.~Sanda, Prog. Theor. Phys. {\bf 110}, 549 (2003).

\bibitem{hfag}
E.~Barberio~{\it et al.}(Heavy Flavor Averaging Group), 0808.1297[hep-ex];
For updates see http://www.slac.standford.edu/xorg/hfag.

\bibitem{npb170}
M.~Merello (CDF Collaboration), \npb(Proc.Suppl.) {\bf 170}, 39 (2007).

\bibitem{pdg2008}
Particle Data Group, C.~Amsler  {\it et al.},\plb {\bf 667}, 1 (2008).

\bibitem{prd52}
I.~Dunietz, \prd {\bf 52}, 3048 (1995).

\bibitem{isg}
I.S.~Gradshteyn and I.M.~Ryzhik, Table of Integrals, Series, and Products,
Academic Press, 1980.

\bibitem{tk07}
T.~Kurimoto, H.N.~Li, and A.I.~Sanda, \prd {\bf 65}, 014007 (2001);
C.D.~Lu and M.Z.~Yang, \epjc {\bf 28}, 515 (2003).

\bibitem{plb555}
H.N.~Li and K.Ukai, \plb{\bf 555}, 197(2003).

\bibitem{epjc695}
H.N.~Li and B.~Melic, \epjc {\bf 11}, 695 (1999).

\end{thebibliography}
\end{document}